\RequirePackage{fix-cm}

\documentclass[english]{article}

\date{}

\usepackage{graphicx}
\usepackage{amsmath}
\usepackage{amssymb}
\usepackage{mathrsfs}
\usepackage{bm}
\usepackage{natbib}
\usepackage{color}
\usepackage{tikz}
\usepackage{eucal}[mathcal]
\usepackage{siunitx}
\usepackage[colorinlistoftodos]{todonotes}
\DeclareMathOperator{\atantwo}{atan2}
\usepackage{caption} 
\allowdisplaybreaks

\begin{document}

\title{A generalization of\\ the equinoctial orbital elements}

\author{Giulio Ba\`u\footnote{University of Pisa, Largo B. Pontecorvo
    5, 56127, Pisa, Italy}\and Javier Hernando-Ayuso\footnote{(in his
    personal capacity) ispace, inc., Sumitomo Fudosan Hamacho
    Building 3F, 3-42-3, Nihonbashi Hamacho, Chuo-ku, Tokyo, Japan
    103-0007} \and Claudio Bombardelli\footnote{Technical University
    of Madrid (UPM), E-28040 Madrid, Spain}}

\maketitle

\begin{abstract}
  We introduce six quantities that generalize the equinoctial orbital
  elements when some or all the perturbing forces that act on the
  propagated body are derived from a disturbing potential. Three of
  the elements define a non-osculating ellipse on the orbital plane,
  other two fix the orientation of the equinoctial reference frame,
  and the last allows one to determine the true longitude of the
  body. The Jacobian matrices of the transformations between the new
  elements and the position and velocity are explicitly given. As a
  possible application we investigate their use in the propagation of
  Earth's artificial satellites showing a remarkable improvement
  compared to the equinoctial orbital elements.
  
\end{abstract}

\section{Introduction}

The set of elements investigated by \citet{BC_1972}:
\begin{equation}
  \begin{aligned}
    a, & & \lambda_0 &= M_0+\omega+\Omega,\\[1ex]
    h &= e\sin(\omega+\Omega), & k &= e\cos(\omega+\Omega),\\[1ex]
    p &= \tan\frac{i}{2}\sin\Omega, & q &= \tan\frac{i}{2}\sin\Omega,
  \end{aligned}
  \label{eq:EqOE}
\end{equation}
where $a$, $e$, $i$, $\Omega$, $\omega$, $M_0$ are the classical
Keplerian elements, are usually recognized as the \emph{equinoctial
  orbital elements}, hereafter EqOE. This expression was coined by
\citet{Arsenault_1970}, who were also the first to introduce the
equinoctial reference frame (see Section \ref{sec:GEqOE_1}). The
appearance of similar elements in Celestial Mechanics dates back to
Lagrange's secular theory of planetary motion. A slight different
version of the quantities $p$, $q$, where the inclination $i$ replaces
$i/2$, is employed in \citet[][p. 130]{Lagrange_1781}. Moreover, the
two quantities $N$ and $M$ introduced at p. 135 of Lagrange's paper,
are a small-inclination approximation of $k$ and $h$, after dividing
by the gravitational parameter.

One of the most relevant variations of the EqOE is due to
\citet{WIO_1985}. They proposed to replace the semi-major axis with
the semi-latus rectum and the mean longitude at epoch ($\lambda_0$)
with the true longitude. In this way, the resulting set can be
applicable to all orbits, while the EqOE work with negative values of
the Keplerian energy only. However, both these sets are singular for
retrograde equatorial orbits (i.e., $i=\pi$), and for rectilinear
motion.

\citet[][Sect. 10.4]{Battin_1999} provided useful relations for the
classical equinoctial elements and their time derivatives, employing
the mean longitude in place of $\lambda_0$. \citet{BC_1972} reported
also the matrix of the partial derivatives of the position and
velocity with respect to the EqOE, and the inverse of that matrix,
along with the Lagrange and Poisson brackets. The authors discuss the
advantages of the EqOE with respect to the universal variables for
computing general perturbations of planets. In a subsequent paper,
\citet{Cefola_1972} focused instead on their use as a special
perturbation method and obtained single-averaged variational equations
in Lagrange's form for different perturbing forces, showing also some
numerical results. Moreover, an alternative set of EqOE was presented
that is non-singular for $i=\pi$ (the singularity is moved to $i=0$).

Thanks to the renewed interest in the EqOE showed in the early 1970s,
they became very appealing for orbit computation programs. For
example, the theory of motion of artificial satellites around the
Earth known as Draper Semianalytic Satellite Theory \citep[see][and
  references therein]{SST_1995}, is based on these
elements. Furthermore, \citet{JAA_1996} showed that orbital elements
can be more effective than Cartesian coordinates in predicting the
shape of uncertainty distributions with the linear error theory,
especially when the observed arc is sufficiently wide. However,
classical orbital elements are strongly affected by nonlinearities
arising from small values of inclination and eccentricity, while
non-singular elements, as the EqOE, are well-suited to the
representation of uncertainties also in these situations
\citep[][pp. 120--121]{MG_2010}. An important advance in this research
field is due to \citet{HAP_2011}, who replaced the semi-major axis
with the mean motion. The resulting \emph{alternate} set of elements
(AEqOE) preserves Gaussianity of the initial state uncertainty through
its propagation at any time in a pure two-body dynamics. This property
was already noticed by \citet[][Sect. 7.4]{MG_2010} for the
\emph{orbit identification} problem. Curiously enough, the mean motion
appears as one of the elements in the forementioned paper by
\citet{Arsenault_1970}.

Generalizations of the EqOE that account for perturbing forces in the
elements definition have been proposed. In a recent work,
\citet{AHA_2021} show the improvement in nonlinear uncertainty
propagation obtained by a set of ``$J_2$ equinoctial orbital elements
(J$_2$EqOE)''. The proposed elements are defined through a multi-step
iterative algorithm that hinges on the Brouwer-Lyddane solution of
the $J_2$-perturbed satellite problem. No direct ordinary differential
equations are provided for the evolution of these elements.
Another relatively recent contribution is due to
\citet{BiriaRussell_2018}, who introduced the \emph{oblate spheroidal}
equinoctial orbital elements, which are formally defined as the
\emph{modified} EqOE of \citet{WIO_1985}, using spheroidal elements
based on Vinti's \citeyearpar{Vinti_1959} theory in place of Keplerian
elements.
The new equinoctial elements have been used by
\citet{BiriaRussell_2020} to write the analytical solution of Vinti's
problem.

In this paper we propose a generalization of the EqOE which is
possible when some or all of the perturbing forces are derivable from
a disturbing potential energy $\mathscr{U}$. In Section
\ref{sec:derivation} we describe how $\mathscr{U}$ can be embedded in
the definitions of the generalized semi-major axis
($\textsl{\textrm{a}}$) and generalized Laplace vector ($\mu{\bf g}$),
which fix a non-osculating ellipse on the orbital plane at every
instant of time. The projections of ${\bf g}$ along the in-plane axes
of the equinoctial reference frame define $p_1$, $p_2$, i.e., the
generalized versions of the elements $h$, $k$. Kepler's equation is
written in a new form where the generalized mean longitude
$\mathcal{L}$ or $\mathcal{L}_0$ appears. The generalized mean motion
$\nu$ and the quantities $q_1$, $q_2$, which coincide with $p$, $q$ in
(\ref{eq:EqOE}), complete our set of \emph{generalized equinoctial
  elements}, hereafter GEqOE. The idea behind the proposed method is
the same that led to the development of two non-singular sets of
orbital elements known as DromoP and EDromo \citep{BauBP_2013,
  BauBPL_2015}. We remark that while DromoP and EDromo employ
redundant variables, the GEqOE consist of only six quantities: $\nu$,
$p_1$, $p_2$, $\mathcal{L}$ (or $\mathcal{L}_0$), $q_1$, $q_2$.

In Sections \ref{sec:pv2ne} -- \ref{sec:FM} we report the
transformation from position and velocity to the GEqOE and its
inverse, the time derivatives of the GEqOE, and the Jacobian matrix of
the transformation together with the inverse of this matrix. In
Section \ref{sec:results} we include some numerical tests to evaluate
the orbit propagation performance of our new elements against the
alternate EqOE as well as the Cartesian elements (i.e., Cowell's
method).

\section{Derivation of the GEqOE}
\label{sec:derivation}
\noindent Consider a point $P$ of mass $m$, which represents a small
body (e.g., a spacecraft), subject to the gravitational attraction of
a body of mass $M$ (e.g., a planet). We introduce a reference frame
\begin{equation}
  \Sigma = \{O;{\bf e}_x,{\bf e}_y, {\bf e}_z\},\label{eq:sigma}
\end{equation}
with the origin in the center of mass $O$ of the planet and fixed
directions in space. Let us use ${\bf r}$ to indicate the position of
$P$ relative to $O$, and $\dot{\bf r}$ for the time derivative of
${\bf r}$ in $\Sigma$. The point mass is also subject to a perturbing
force ${\bf F}$:
\begin{equation}
  {\bf F} = {\bf P} - \nabla\mathscr{U}({\bf r},t),
  \label{eq:F}
\end{equation}
where $\mathscr{U}$ is the opposite of the disturbing potential, and
can depend on ${\bf r}$ and time $t$. For future use, we introduce the
\emph{orbital} reference frame $\Sigma_{\rm or}=\{O;{\bf e}_r,{\bf
  e}_f,{\bf e}_h\}$, where
\begin{equation*}
  {\bf e}_r = \frac{{\bf r}}{|{\bf r}|},\quad
  {\bf e}_f = {\bf e}_h\times{\bf e}_r,\quad
  {\bf e}_h = \frac{{\bf r}\times\dot{\bf r}}{|{\bf r}\times\dot{\bf r}|}.
\end{equation*}

In the remainder of the article, we will refer to the
\emph{equinoctial orbital elements} to indicate the set of elements
presented in \citet{BC_1972}.

\subsection{The non-osculating ellipse $\Gamma$}
\label{sec:non_osc}
\begin{figure*}
  \centering
  \includegraphics[width=0.95\textwidth]{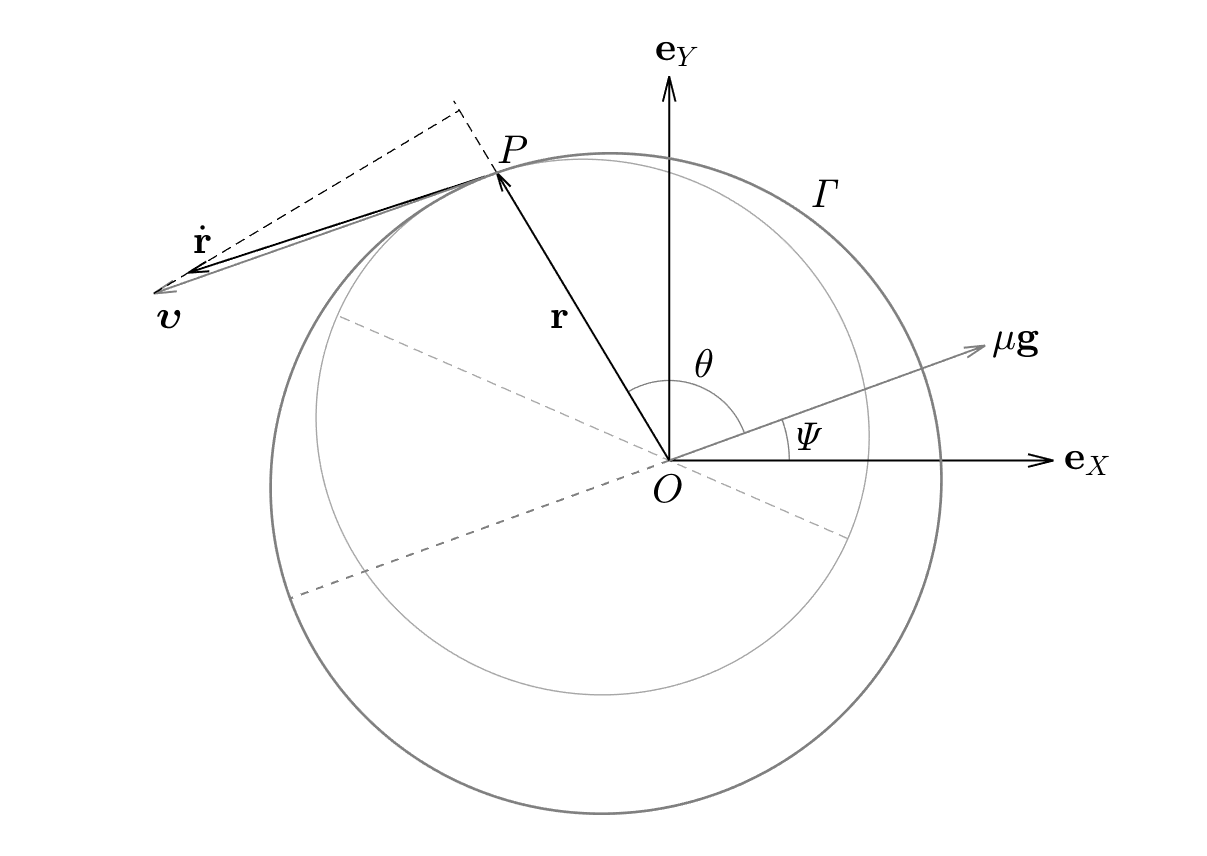}
  \caption{View from the angular momentum vector of the
    \emph{osculating} ellipse (in light grey) and the
    \emph{non-osculating} ellipse $\Gamma$ described in Section
    (\ref{sec:non_osc}). The generalized true anomaly $\theta$ and
    generalized longitude of pericenter $\Psi$ are also shown. The
    velocities $\dot{\bf r}$ and $\bm\upsilon$ of $P$ along the two
    ellipses have the same radial component $\dot{r}$.}
  \label{fig:non_osc}
\end{figure*}

\noindent Let $h=|{\bf r}\times\dot{\bf r}|$ be the magnitude of the
angular momentum vector of $P$ and $r=|{\bf r}|$ the orbital
distance. Assume that $r$ and $h$ are strictly positive quantities. We
define the \emph{effective} potential energy as
\[
  \mathscr{U}_{\textsf{eff}}({\bf r},\dot{\bf r},t) = \frac{h^2}{2r^2}+\mathscr{U}({\bf r},t).
\]
Then, the total energy $\mathscr{E}$ can be written in the form
\[
\mathscr{E}({\bf r},\dot{\bf r},t) = \frac{1}{2}\dot{r}^2-\frac{\mu}{r}+
\mathscr{U}_{\textsf{eff}}({\bf r},\dot{\bf r},t),
\]
where $\dot{r}$ is the radial velocity and $\mu=G(M+m)$, with $G$ the
gravitational constant. We introduce the generalized angular momentum
\begin{equation}
  c = \sqrt{2r^2\mathscr{U}_{\textsf{eff}}},
  \label{eq:c}
\end{equation}
and the generalized velocity vector
\begin{equation}
  \bm\upsilon = \dot{r}\,{\bf e}_r+\frac{c}{r}{\bf e}_f.
  \label{eq:gv}
\end{equation}
The pair of vectors $({\bf r}, {\bm\upsilon})$ defines a
\emph{non-osculating} ellipse $\Gamma$, having one focus located at
the center of mass of the primary body of attraction. Its shape is
fixed by the generalized semi-major axis and generalized eccentricity,
given by
\begin{align}
  \textsl{\textrm{a}} &= -\frac{\mu}{2\mathscr{E}},\label{eq:ga}\\[1ex]
  g &= \frac{1}{\mu}\sqrt{\mu^2+2\mathscr{E}c^2}.\label{eq:ge}
\end{align}
Denoting by $e$ the eccentricity and by $\mathscr{E}_K$ the Keplerian
energy, we find
\[
g^2 = e^2 + \frac{2\mathscr{U}}{\mu^2}\left[h^2+2r^2(\mathscr{E}_K+\mathscr{U})\right].
\]
The ellipse $\Gamma$ lies on the orbital plane and its orientation
on this plane is fixed by the generalized Laplace vector (see Figure
\ref{fig:non_osc})
\[
\mu{\bf g} = {\bm\upsilon}\times({\bf r}\times{\bm\upsilon})-\mu{\bf e}_r,
\]
where $|{\bf g}|=g$.

\emph{Remark 1} Note that when $\dot{r}=0$, $P$ is at the
pericenter/apocenter of both the osculating conic defined by the
Keplerian orbital elements and the non-osculating ellipse $\Gamma$.

Let us introduce the generalized true anomaly $\theta$ through the relations
\begin{align}
  g\cos\theta & = \frac{c^2}{\mu r}-1,\label{eq:cth}\\[1ex]
  g\sin\theta & = \frac{c\dot{r}}{\mu},\label{eq:sth}
\end{align}
which are analogous to the well-known relations for the Kepler problem
\begin{align*}
  e\cos f & = \frac{h^2}{\mu r}-1,\\[1ex]
  e\sin f & = \frac{h\dot{r}}{\mu},
\end{align*}
where $f$ is the true anomaly. The angle $\theta$ allows us to recover
the orientation of the radial direction from that of ${\bf g}$.

\subsection{The elements $\nu$, $p_1$, $p_2$, $\mathcal{L}$}
\label{sec:GEqOE_1}
\noindent Consider the classical equinoctial reference frame
\[
\Sigma_{\textsf{eq}} = \{O;{\bf e}_X,{\bf e}_Y,{\bf e}_Z\}.
\]
The axis $OX$, associated to ${\bf e}_X$, is rotated of $\Omega$ from
the ascending node in the retrograde direction; the axis $OY$,
associated to ${\bf e}_Y$, is rotated of $\pi/2$ from the axis $OX$ in
the direction of the motion. The unit vector ${\bf e}_Z$ completes the
right-handed orthonormal basis. The angular displacement between the
direction of ${\bf e}_r$ and the \emph{departure} direction defined
by ${\bf e}_X$ is called true longitude, and is given by
\begin{equation}
  L = \varpi+f,\label{eq:tl}
\end{equation}
where
\[
  \varpi = \omega+\Omega,
\]
with $\omega$ the argument of pericenter and $\Omega$ the longitude
of the ascending node. We define the angular variable $\Psi$ as
\begin{equation}
  \Psi = L-\theta.\label{eq:Psi}
\end{equation}
When $\mathscr{U}=0$ the angle $\theta$ coincides with $f$ and thereby
$\Psi=\varpi$, which becomes constant if we have also ${\bf P}={\bf
  0}$. It is straightforward to check that through the angle $\Psi$ we
can obtain the direction of the Laplace vector from the direction of
${\bf e}_X$, and for this reason we call it the generalized longitude
of pericenter (see Figure \ref{fig:non_osc}).

The first three elements of the new set are defined as
\begin{align}
  \nu & := \frac{1}{\mu}(-2\mathscr{E})^{3/2},\label{eq:enne}\\[1ex]
  p_1 & := g\sin\Psi,\label{eq:p1}\\[1ex]
  p_2 & := g\cos\Psi\label{eq:p2},
\end{align}
where $\nu$ is the generalized form of the mean motion $n$, and
$p_1$, $p_2$ are the generalized versions of the equinoctial orbital
elements $h$, $k$ \citep{BC_1972}. For later use, we introduce
the generalized semi-latus rectum
\begin{equation} 
  \varrho = \textsl{\textrm{a}}(1-g^2),\label{eq:gsr}
\end{equation}
and note that the following formula holds:
\begin{equation}
  c^2=\mu\varrho,\label{eq:c2_1}
\end{equation}
which is obtained from (\ref{eq:ga}), (\ref{eq:ge}),
(\ref{eq:gsr}). Moreover, since
\begin{align}
  \textsl{\textrm{a}} & = \biggl(\frac{\mu}{\nu^2}\biggr)^{1/3},\label{eq:ga_2}\\[1ex]
  g^2 & = p_1^2+p_2^2,\nonumber
\end{align}
we can write $c$ as function of $\nu$, $p_1$, $p_2$:
\begin{equation}
  c=\biggl(\frac{\mu^2}{\nu}\biggr)^{1/3}\sqrt{1-
  p_1^2-p_2^2}.\label{eq:c_2}
\end{equation}

At this point we need to make another step to define the fourth
generalized equinoctial element. We first introduce the generalized
eccentric anomaly $G$ through the relations
\begin{align}
  r & = \textsl{\textrm{a}}(1 - g\cos G),\label{eq:r}\\[1ex]
  r\dot{r} & = g\sqrt{\mu\textsl{\textrm{a}}}\sin G,\label{eq:rrd}
\end{align}
which are analogous to the well-known relations for the Kepler problem
\begin{align*}
  r & = a(1 - e\cos E),\\[1ex]
  r\dot{r} & = e\sqrt{\mu a}\sin E,
\end{align*}
where $a$, $E$ are the semi-major axis and eccentric anomaly,
respectively. Then, the generalized Kepler's equation can be written as
\begin{equation}
  \mathcal{M} = G-g\sin G,\label{eq:KE}
\end{equation}
where $\mathcal{M}$ is the generalized mean anomaly
\begin{equation}
  \mathcal{M} = \nu(t-t_0),\label{eq:emme}
\end{equation}
and $t_0$ is the time of passage through the pericenter of the ellipse
$\Gamma$ (see Remark 1).

We include in the GEqOE the generalized mean longitude
\begin{equation}
  \mathcal{L} := \mathcal{M}+\Psi.\label{eq:elle}
\end{equation}
After defining in a similar way the generalized eccentric longitude as
\begin{equation}
  \mathcal{K} = G+\Psi,\label{eq:K}
\end{equation}
we can put equation (\ref{eq:emme}) in the form
\begin{equation}
  \mathcal{L} = \mathcal{K}+p_1\cos\mathcal{K}-p_2\sin\mathcal{K},\label{eq:elle_2}
\end{equation}
where the right-hand side is derived from (\ref{eq:KE}) by taking into
account (\ref{eq:p1}), (\ref{eq:p2}), (\ref{eq:elle}),
(\ref{eq:K}). If we know the values of $p_1$, $p_2$, $\mathcal{L}$ we can
compute $\mathcal{K}$ by solving Kepler's equation
(\ref{eq:elle_2}). The orbital distance and the radial velocity are
obtained by means of the formulae:
\begin{align}
  r & = \textsl{\textrm{a}}(1-p_1\sin\mathcal{K}-p_2\cos\mathcal{K}),
  \label{eq:r_2}\\[1ex]
  \dot{r} & = \frac{\sqrt{\mu\textsl{\textrm{a}}}}{r}(p_2\sin\mathcal{K}-
  p_1\cos\mathcal{K}),\label{eq:rrd_2}
\end{align}
which follow from equations (\ref{eq:r}), (\ref{eq:rrd}) where we use
the definitions (\ref{eq:p1}), (\ref{eq:p2}),
(\ref{eq:K}). Considering also relation (\ref{eq:ga_2}), we recognize
that $r$ and $\dot{r}$ are known from the first four GEqOE, i.e.,
$\nu$, $p_1$, $p_2$, $\mathcal{L}$, which are defined in equations
(\ref{eq:enne}), (\ref{eq:p1}), (\ref{eq:p2}), (\ref{eq:elle}).

It is worth noting that $r$ and $\dot{r}$ can also be expressed as
functions of the true longitude. From equations (\ref{eq:cth}),
(\ref{eq:sth}) and (\ref{eq:Psi}), (\ref{eq:p1}), (\ref{eq:p2}) we
have
\begin{align}
  r & = \frac{\varrho}{1+p_1\sin L+p_2\cos L},\label{eq:r_3}\\[1ex]
  \dot{r} & = \frac{\mu}{c}(p_2\sin L-p_1\cos L),\label{eq:rrd_3}  
\end{align}
where $\varrho$ is introduced in (\ref{eq:gsr}).

\subsection{The remaining elements}
\noindent The three elements $\nu$, $p_1$, $p_2$ determine the shape
and orientation of the non-osculating ellipse $\Gamma$ on the orbital
plane, and $\mathcal{L}$ fixes the position of $P$ with respect to
$\Sigma_{\textsf{eq}}$. Therefore, the remaining elements of the
proposed set need to characterize the orientation of
$\Sigma_{\textsf{eq}}$ with respect to $\Sigma$ (see equation
\ref{eq:sigma}), which can be recovered by applying the sequence of
rotations $\Omega$, $i$, $-\Omega$, where $i$ is the orbital
inclination.

The two elements $p$, $q$ in \citet{BC_1972}, that is:
\begin{align}
  q_1 & := \tan\frac{i}{2}\sin\Omega,\label{eq:q1}\\[1ex]
  q_2 & := \tan\frac{i}{2}\cos\Omega\label{eq:q2}
\end{align}
satisfy our request, and therefore it is natural to include them in
the set of GEqOE. An alternative to $q_1$, $q_2$ is represented by the
Euler parameters $e_1$, $e_2$, $e_3$ that define the orientation of
$\Sigma_{\textsf{eq}}$ with respect to $\Sigma$
\citep[][p. 155]{Goldstein_1980}\footnote{One of the Euler parameters
  is identically equal to 0.}:
\begin{equation}
  e_1 := \cos\frac{i}{2}\cos\Omega,\quad e_2 := \sin\frac{i}{2},\quad
  e_3 := \cos\frac{i}{2}\sin\Omega.
  \label{eq:e123}
\end{equation}
Note that both these elements and $q_1$, $q_2$ suffer of the
singularity for $i=\pi$. The Euler parameters allow us to partially
control the error accumulation during the propagation by monitoring
the quantity $e_1^2+e_2^2+e_3^2$. On the other hand, they make the set
of GEqOE redundant, increasing the dimension of the state vector from
6 to 7. We refer to Section \ref{sec:EulPar} for more details about
the alternative formulation with $e_1$, $e_2$, $e_3$ in place of
$q_1$, $q_2$.

\subsection{Summary}
\label{sec:summary}
\noindent A set of \emph{generalized equinoctial orbital elements}
consists of
\[
\nu\,\,({\rm eq.}~\ref{eq:enne}),
\quad p_1\,\,({\rm eq.}~\ref{eq:p1}),\quad
p_2\,\,({\rm eq.}~\ref{eq:p2}),\quad
\mathcal{L}\,\,({\rm eq.}~\ref{eq:elle}),
\]
along with
\[
q_1\,\,({\rm eq.}~\ref{eq:q1}),
\quad q_2\,\,({\rm eq.}~\ref{eq:q2}).
\]
The generalized mean longitude $\mathcal{L}$ can be replaced by the
generalized mean longitude at epoch $\mathcal{L}_0$ as shown in
Section \ref{sec:cte}. These sets of elements represent two
generalizations of the alternate equinoctial orbital elements proposed
by \citet{HAP_2011} (see the Introduction), with an improved
propagation performance, as it will be shown in Section
\ref{sec:results}.

\section{From position and velocity to the GEqOE}
\label{sec:pv2ne}
\noindent Assume that we know the position (${\bf r}$) and velocity
($\dot{\bf r}$) at some time $t$ with respect to the reference frame
$\Sigma$ (see equation \ref{eq:sigma}). We want to determine the
values of the new elements.

First, we get the quantities
\[
r=|{\bf r}|,\qquad\dot{r}=\frac{{\bf r}\cdot\dot{\bf r}}{r},
\]
and compute the total energy:
\[
  \mathscr{E}({\bf r},\dot{\bf r},t) = \mathscr{E}_K({\bf r},\dot{\bf r})+\mathscr{U}({\bf r},t),
\]
where $\mathscr{E}_K$ is the Keplerian energy and the disturbing
potential energy $\mathscr{U}$ does not depend on $\dot{\bf r}$. The
element $\nu$ is obtained from equation (\ref{eq:enne}).

From the Keplerian orbital elements $\Omega$, $i$, which are
determined by classical formulae, we compute $q_1$, $q_2$ through
equations (\ref{eq:q1}), (\ref{eq:q2}), and the unit vectors ${\bf
  e}_X$, ${\bf e}_Y$ of the equinoctial reference frame
$\Sigma_{\textsf{eq}}$ by
\begin{equation}
  \begin{split}
    {\bf e}_X & = \frac{1}{1+q_1^2+q_2^2}\left(1-q_1^2+q_2^2,\,\,
    2q_1q_2,\,\,-2q_1\right)^T,\\[1ex]
    {\bf e}_Y & = \frac{1}{1+q_1^2+q_2^2}\left(2q_1q_2,\,\,
    1+q_1^2-q_2^2,\,\,2q_2\right)^T.
  \end{split}
  \label{eq:eXY}
\end{equation}

Inversion of relations (\ref{eq:r_3}), (\ref{eq:rrd_3}) yields
\begin{align}
  p_1 &= \Bigl(\frac{\varrho}{r}-1\Bigr)\sin L-\frac{c\dot{r}}{\mu}\cos L,
  \label{eq:p1_L}\\[1ex]
  p_2 &= \Bigl(\frac{\varrho}{r}-1\Bigr)\cos L+\frac{c\dot{r}}{\mu}\sin L,
  \label{eq:p2_L}
\end{align}
where
\[
\cos L = {{\bf e}_r}\cdot{{\bf e}_X},\qquad
\sin L = {{\bf e}_r}\cdot{{\bf e}_Y},
\]
with ${\bf e}_r$ the radial unit vector, and ${\bf e}_X$, ${\bf e}_Y$
given by (\ref{eq:eXY}). The quantity $c$ is obtained using formula
(\ref{eq:c}), wherein $h=|{\bf r}\times\dot{\bf r}|$.

In order to determine the generalized mean longitude $\mathcal{L}$, we need
to know the value of the generalized eccentric longitude
$\mathcal{K}$. Let us introduce
\begin{align*}
  S & = (\mu+c\textsl{\textrm{w}}-r\dot{r}^2)\sin L-
      \dot{r}(c+\textsl{\textrm{w}}r)\cos L,\\[1ex]
  C & = (\mu+c\textsl{\textrm{w}}-r\dot{r}^2)\cos L+
      \dot{r}(c+\textsl{\textrm{w}}r)\sin L.
\end{align*}
It is possible to show that (see Appendix \ref{sec:sKcK})
\begin{equation}
  \sin\mathcal{K} = \frac{S}{\mu+c\textsl{\textrm{w}}},\quad
  \cos\mathcal{K} = \frac{C}{\mu+c\textsl{\textrm{w}}},\label{eq:sKcK}
\end{equation}
where
\begin{equation}
  \textsl{\textrm{w}} = \sqrt{\frac{\mu}{\textsl{\textrm{a}}}},
  \label{eq:w}
\end{equation}
and $\textsl{\textrm{a}}$ depends on $\nu$ (see \ref{eq:ga_2}). The value of
$\mathcal{L}$ is found by means of the generalized Kepler's equation
(\ref{eq:elle_2}), which is written as
\[
\mathcal{L} = \atantwo(S,C)+\frac{Cp_1-Sp_2}{\mu+c\textsl{\textrm{w}}}.
\]

Finally, we point out that an alternative way to get $p_1$, $p_2$
is given by the formulae:
\begin{align*}
  p_1 & = \frac{1}{\mu+c\textsl{\textrm{w}}}\left[\Bigl(1-\frac{r}{\textsl{\textrm{a}}}\Bigr)S-
  \frac{r\dot{r}}{\sqrt{\mu\textsl{\textrm{a}}}}C\right],\\[1ex]
  p_2 & = \frac{1}{\mu+c\textsl{\textrm{w}}}\left[\frac{r\dot{r}}{\sqrt{\mu\textsl{\textrm{a}}}}S+
  \Bigl(1-\frac{r}{\textsl{\textrm{a}}}\Bigr)C\right],
\end{align*}
which are derived by solving equations (\ref{eq:r_2}),
(\ref{eq:rrd_2}) for $p_1$, $p_2$ and making the substitutions in
(\ref{eq:sKcK}).

\section{From the GEqOE to position and velocity}
\label{sec:ne2pv}
\noindent Assume that we know the values taken by the new elements at
some time $t$ and we want to find ${\bf r}$ and $\dot{\bf r}$ at that
epoch.

We first solve Kepler's equation (\ref{eq:elle_2}) for
$\mathcal{K}$. Then, we compute $\textsl{\textrm{a}}$ from
(\ref{eq:ga_2}), and obtain $r$ and $\dot{r}$ from (\ref{eq:r_2}) and
(\ref{eq:rrd_2}). By combining these two equations with
(\ref{eq:r_3}), (\ref{eq:rrd_3}), and considering (\ref{eq:gsr}),
(\ref{eq:c2_1}), we find
\begin{equation}
  \begin{split}
    \sin L & = \frac{\textsl{\textrm{a}}}{r}\left[\alpha p_1p_2\cos\mathcal{K}+
    (1-\alpha p_2^2)\sin\mathcal{K}-p_1\right],\\[1ex]
    \cos L & = \frac{\textsl{\textrm{a}}}{r}\left[\alpha p_1p_2\sin\mathcal{K}+
    (1-\alpha p_1^2)\cos\mathcal{K}-p_2\right],
  \end{split}
  \label{eq:scL}
\end{equation}
where
\begin{equation}
  \alpha = \frac{1}{1+\sqrt{1-p_1^2-p_2^2}}.
  \label{eq:alpha}
\end{equation}
After computing the unit vectors ${\bf e}_X$, ${\bf e}_Y$ of the
equinoctial reference frame by means of (\ref{eq:eXY}), we can obtain
the unit vectors ${\bf e}_r$, ${\bf e}_f$ of the orbital
basis through the rotation:
\begin{equation}
  \begin{split}
  {\bf e}_r & = {\bf e}_X\cos L+{\bf e}_Y\sin L,\\[1ex]
  {\bf e}_f & = {\bf e}_Y\cos L-{\bf e}_X\sin L.
  \end{split}
  \label{eq:erhotheta}
\end{equation}

Finally, the position and velocity vectors are given by the formulae
\begin{equation}
  {\bf r} = r{\bf e}_r,\qquad\dot{\bf r} = \dot{r}{\bf e}_r
  +\frac{h}{r}{\bf e}_f.
  \label{eq:r_rd}
\end{equation}
Since the function $\mathscr{U}$ does not depend of $\dot{\bf r}$ we
can use equation (\ref{eq:c}) to calculate $h$:
\[
h=\sqrt{c^2-2r^2\mathscr{U}({\bf r},t)},
\]
where $c^2$ is obtained from (\ref{eq:c_2}).

\section{Time derivatives of the GEqOE}
\label{sec:dpdt}
\noindent The time derivative of $\nu$ is
\begin{equation}
  \dot{\nu} = -3\biggl(\frac{\nu}{\mu^2}\biggr)^{\!1/3}\dot{\mathscr{E}},
  \label{eq:dnu}
\end{equation}
where
\begin{equation}
  \dot{\mathscr{E}} = \mathscr{U}_t+{\bf P}\cdot\dot{\bf r} = \mathscr{U}_t+
  \dot{r}P_r+\frac{h}{r}P_f,\label{eq:dE}
\end{equation}
and $\mathscr{U}_t$ denotes the partial derivative of
$\mathscr{U}({\bf r},t)$ with respect to $t$.

The angular velocity of $\Sigma_{\textsf{eq}}$ with respect to
$\Sigma$ is
\[
{\bf w} = w_X{\bf e}_X + w_Y{\bf e}_Y + w_h{\bf e}_h,
\]
where
\begin{align*}
  w_X & = F_h\frac{r}{h}\cos L,\\[1ex]
  w_Y & = F_h\frac{r}{h}\sin L,\\[1ex]
  w_h & = \frac{h}{r^2}-\dot{L}=-F_h\frac{r}{h}\tan{\frac{i}{2}}\sin(\omega+f),
\end{align*}
and $F_h$ is the projection of the perturbing force along ${\bf
  e}_h$. Note that $w_h$ is not defined when $i=\pi$. If we divide
$w_X$, $w_Y$, $w_h$ by $F_hr/h$, and denote the resulting quantities
with $\hat{w}_X$, $\hat{w}_Y$, $\hat{w}_h$, we have
\begin{equation}
  \begin{split}
    \hat{w}_X &= \cos L,\\
    \hat{w}_Y &= \sin L,\\
    \hat{w}_h &= q_1\cos L-q_2\sin L,
  \end{split}
  \label{eq:w_XYh}
\end{equation}
where $q_1$, $q_2$ are defined in (\ref{eq:q1}), (\ref{eq:q2}). The
expression for $\hat{w}_h$ is obtained noting that
\[
\omega+f = L-\Omega.
\]

For the time derivatives of $p_1$, $p_2$ we find (see the derivation
in Appendix \ref{sec:dp12dt})
\begin{align}
  \dot{p}_1 & = p_2\biggl(\frac{h-c}{r^2}-\frac{r}{h}\hat{w}_hF_h\biggr)
  +\frac{1}{c}\Bigl(\frac{r\dot{r}}{c}p_1+\tilde{\varsigma}p_2+\varsigma\cos L
  \Bigr)(2\mathscr{U}-rF_r)\nonumber\\[0.5ex]
  & \quad\, +\frac{r}{\mu}(\varsigma\,p_1+\tilde{\varsigma}\sin L)
  \dot{\mathscr{E}},\label{eq:dp1}\\[1ex]
  \dot{p}_2 & = p_1\biggl(\frac{r}{h}\hat{w}_hF_h-\frac{h-c}{r^2}\biggr)
  +\frac{1}{c}\Bigl(\frac{r\dot{r}}{c}p_2-\tilde{\varsigma}p_1-\varsigma\sin L
  \Bigr)(2\mathscr{U}-rF_r)\nonumber\\[0.5ex]
  & \quad\, +\frac{r}{\mu}(\varsigma\,p_2+\tilde{\varsigma}\cos L)
  \dot{\mathscr{E}},\label{eq:dp2}
\end{align}
where $F_r={\bf F}\cdot{\bf e}_r$, $F_h={\bf F}\cdot{\bf
  e}_h$ and we use the non-dimensional quantities
\begin{equation}
  \varsigma = \frac{r}{\varrho},\qquad\tilde{\varsigma} = 1+\varsigma.\label{eq:aux}
\end{equation}
Concerning the element $\mathcal{L}$, we can write (see the
derivation in Appendix \ref{sec:delldt})
\begin{equation}
  \begin{split}
  \dot{\mathcal{L}} &= \nu+\frac{h-c}{r^2}-\frac{r}{h}
  \hat{w}_hF_h+\Bigl(\frac{r\dot{r}c}{\mu^2}\tilde{\varsigma}\alpha\Bigr)\dot{\mathscr{E}}\\
  & \quad\,+\frac{1}{c}\biggl[\frac{1}{\alpha}+\alpha\Bigl(1-\frac{r}{\textsl{\textrm{a}}}\Bigr)\biggr]
  (2\mathscr{U}-rF_r),
  \end{split}
  \label{eq:dell}
\end{equation}
where $\alpha$ is defined in (\ref{eq:alpha}). For the remaining two
elements $q_1$, $q_2$ we need the derivatives \citep[see][eqs. 10.51,
  10.52, p. 493]{Battin_1999}
\begin{align}
  \dot{q}_1 &= \frac{r}{2h}F_h(1+q_1^2+q_2^2)\sin L,\label{eq:dq1}\\[1ex]
  \dot{q}_2 &= \frac{r}{2h}F_h(1+q_1^2+q_2^2)\cos L.\label{eq:dq2}
\end{align}

The right-hand side of equations (\ref{eq:dnu}), (\ref{eq:dp1}),
(\ref{eq:dp2}), (\ref{eq:dell}), (\ref{eq:dq1}), (\ref{eq:dq2}) can be
efficiently computed by following the procedure outlined in Section
\ref{sec:ne2pv}.

\subsection{Constant time element}
\label{sec:cte}
\noindent The equinoctial elements presented in \citet{BC_1972}
comprise the mean longitude at epoch $\lambda_0=\varpi+nt_0$, where we
recall that $n$ is the mean motion and $t_0$ the time of pericenter
passage. This quantity is a constant of the motion when the
perturbations are turned off, and being related to the physical time
we can refer to $\lambda_0$ as a \emph{constant} time element. On the
other hand, the generalized mean longitude $\mathcal{L}$ included in
the GEqOE varies linearly with time along Keplerian motion (see
equation \ref{eq:dell}), and therefore it is a \emph{linear} time
element.

In place of $\mathcal{L}$, we may consider the generalized mean motion
at epoch $\mathcal{L}_0$, which we define as\footnote{Another possible
  definition is $\tilde{\mathcal{L}}_0 := \Psi+\nu t_0$, which
  represents a direct generalization of the element $\lambda_0$. In
  this case we have $\tilde{\mathcal{L}}_0=\mathcal{L}-\mathcal{M}+\nu
  t_0$.\label{f:L0}}
\[
\mathcal{L}_0 := \Psi-\nu t_0.
\]
Using equations (\ref{eq:emme}), (\ref{eq:elle}), we see that
\[
\mathcal{L}_0 = \mathcal{L}-\nu t,
\]
and therefore its time derivative can be computed from
(\ref{eq:dnu}), (\ref{eq:dell}), resulting in
\begin{align*}
  \dot{\mathcal{L}}_0 & = \frac{h-c}{r^2}-\frac{r}{h}\hat{w}_hF_h+
  \biggl[3t\biggl(\frac{\nu}{\mu}\biggr)^{\!1/3}+\frac{r\dot{r}c}{\mu^2}
  \tilde{\varsigma}\alpha\biggr]\dot{\mathscr{E}}\\ &
  \quad\,+\frac{1}{c}\biggl[\frac{1}{\alpha}+\alpha\Bigl(1-\frac{r}{\textsl{\textrm{a}}}\Bigr)\biggr]
  (2\mathscr{U}-rF_r),
\end{align*}
where $\alpha$, $\tilde{\varsigma}$ are introduced in
(\ref{eq:alpha}), (\ref{eq:aux}). We note that a term dependent
explicitly on time arises in the expression of $\dot{\mathcal{L}}_0$, which
is not present in $\dot{\mathcal{L}}$. For long-term propagations this term
may grow enough to deteriorate the efficiency of the propagation.

\subsection{Alternative formulation}
\label{sec:EulPar}
If the Euler parameters $e_1$, $e_2$, $e_3$ given in (\ref{eq:e123})
are used in place of $q_1$, $q_2$, then, ${\bf e}_X$, ${\bf e}_Y$ are
computed from the formulae
\begin{align*}
  {\bf e}_X & = \left(e_1^2+e_2^2-e_3^2,\,\,
  -2e_1e_3,\,\,2e_2e_3\right)^T,\\[1ex]
  {\bf e}_Y & = \left(2e_1e_3,\,\,e_1^2-e_2^2-e_3^2,\,\,
  -2e_1e_2\right)^T.
\end{align*}
The time derivatives of the Euler parameters read
\begin{align*}
  \dot{e}_1 & =\frac{r}{2h}F_h(\hat{w}_he_2+\hat{w}_Xe_3),\\[1ex]
  \dot{e}_2 & =-\frac{r}{2h}F_h(\hat{w}_he_1-\hat{w}_Ye_3),\\[1ex]
  \dot{e}_3 & =-\frac{r}{2h}F_h(\hat{w}_Xe_1+\hat{w}_Ye_2),
\end{align*}
where $\hat{w}_X$, $\hat{w}_Y$ are defined in (\ref{eq:w_XYh}) and
\[
\hat{w}_h = \frac{e_2}{e_1^2+e_3^2}(e_3\cos L-e_1\sin L),
\]
with $\sin L$, $\cos L$ obtained by (\ref{eq:scL}).

\section{The fundamental matrix and its inverse}
\label{sec:FM}
\noindent The fundamental matrix is defined as the matrix of the
partial derivatives of position and velocity with respect to the set of
elements used for describing the motion \citep{B_1970}. We consider in
this section the two sets of GEqOE given by $\nu$, $p_1$, $p_2$,
$q_1$, $q_2$, along with either $\mathcal{L}$ or $\mathcal{L}_0$. In
\citet{BC_1972}, the fundamental matrix and its inverse are expressed
using the perifocal reference frame. However, since its basis is not
defined when the eccentricity is zero, we prefer to have the unit
vectors of the equinoctial and orbital reference frames appearing
directly in (\ref{eq:dr_dp}), (\ref{eq:ddr_dp}), (\ref{eq:dp_dr}),
(\ref{eq:dp_ddr}) \citep[see][]{SST_1995}. We use the notation:
\begin{align*}
  X &= r\cos L, & Y &= r\sin L,\\[1ex]
  \dot{X} &= \dot{r}\cos L-\frac{h}{r}\sin L, &
  \dot{Y} &= \dot{r}\sin L+\frac{h}{r}\cos L,
\end{align*}
for the components of the position and velocity vectors along the
directions of the unit vectors ${\bf e}_X$, ${\bf e}_Y$, that is
\[
  {\bf r} = X{\bf e}_X+Y{\bf e}_Y,\qquad
  \dot{\bf r} = \dot{X}{\bf e}_X+\dot{Y}{\bf e}_Y.
\]
Moreover, we introduce the non-dimensional quantities
\begin{align}
  \beta &= \sqrt{1-p_1^2-p_2^2}, & \gamma &= 1+q_1^2+q_2^2,
  \label{eq:bega}
\end{align}
where $\beta=0$, $\gamma=0$ when $g=0$, $i=\pi$, respectively.

\subsection{Partial derivatives of position and velocity with respect to the GEqOE}

We obtain the partial derivatives of ${\bf r}$, $\dot{\bf r}$ with
respect to the GEqOE by direct differentiation of equations
(\ref{eq:r_rd}), wherein $r$, $\dot{r}$ and ${\bf e}_r$, ${\bf e}_f$
are replaced by the expressions reported in (\ref{eq:r_2}),
(\ref{eq:rrd_2}) and (\ref{eq:erhotheta}), respectively. Relations
(\ref{eq:elle_2}), (\ref{eq:scL}) are also necessary. Regarding the
position, we have:
\begin{equation}
  \begin{split}
    \frac{\partial{\bf r}}{\partial \nu} &= -\frac{2}{3\nu}{\bf r},\\[1ex]
    \frac{\partial{\bf r}}{\partial p_1} &= -\textsl{\textrm{a}}\Bigl(\frac{\alpha p_2}{\beta}{\bf q}+
    {\bf e}_Y\Bigr)-\frac{\textsl{\textrm{a}}}{\varrho}\left[X+p_2(r+\alpha\beta\textsl{\textrm{a}})
    \right]{\bf e}_f,
    \\[1ex]
    \frac{\partial{\bf r}}{\partial p_2} &= \textsl{\textrm{a}}\Bigl(\frac{\alpha p_1}{\beta}{\bf q}-
    {\bf e}_X\Bigr)+\frac{\textsl{\textrm{a}}}{\varrho}\left[Y+p_1(r+\alpha\beta\textsl{\textrm{a}})
    \right]{\bf e}_f,
    \\[1ex]
    \frac{\partial{\bf r}}{\partial \mathcal{L}} &= \frac{1}{\nu}{\bm\upsilon},\\[1ex]
    \frac{\partial{\bf r}}{\partial q_1} &= -\frac{2}{\gamma}
    (rq_2{\bf e}_f+X{\bf e}_h),\\[1ex]
    \frac{\partial{\bf r}}{\partial q_2} &= \frac{2}{\gamma}
    (rq_1{\bf e}_f+Y{\bf e}_h),
  \end{split}
  \label{eq:dr_dp}
\end{equation}
where the generalized velocity $\bm\upsilon$ and the non-dimensional
quantity $\alpha$ are introduced in (\ref{eq:gv}), (\ref{eq:alpha}),
respectively, and
\[
{\bf q} = p_2{\bf e}_Y-p_1{\bf e}_X.
\]

\emph{Remark 2.} It is possible to prove that the derivatives of ${\bf
  r}$ with respect to $\textsl{\textrm{a}}$, $p_1$, $p_2$,
$\tilde{\mathcal{L}}_0$ (see the footnote \ref{f:L0}), $q_1$, $q_2$
can be written in the same form as the derivatives of ${\bf r}$ with
respect to $a$, $h$, $k$, $\lambda_0$, $p$, $q$ that are reported in
\citet[][Table I]{BC_1972}. We just have to replace in the latters the
osculating eccentric anomaly ($E$), eccentricity ($e$), semi-major
axis ($a$), longitude of pericenter ($\omega+\Omega$) by $G$, $g$,
$\textsl{\textrm{a}}$, $\Psi$, respectively, and the unit vectors
${\bf e}_p$, ${\bf e}_q$ of the perifocal reference
frame\footnote{These two unit vectors are denoted by ${\bf P}$, ${\bf
    Q}$ in \citet{BC_1972}.} by their generalized versions ${\bf
  e}_p'$, ${\bf e}_q'$, which are defined as (see also Figure
\ref{fig:perifocal})
\begin{align*}
  {\bf e}_p' &= {\bf e}_X\cos\Psi+{\bf e}_Y\sin\Psi,\\[1ex]
  {\bf e}_q' &= {\bf e}_Y\cos\Psi-{\bf e}_X\sin\Psi,
\end{align*}
where $\Psi$ is given in (\ref{eq:Psi}).

The partial derivatives of the velocity with respect to the GEqOE are:
\begin{equation}
  \begin{split}
    \frac{\partial{\dot{\bf r}}}{\partial \nu} &= \frac{1}{3\nu}\dot{\bf r}+f_0{\bf e}_f,\\[1ex]
    \frac{\partial{\dot{\bf r}}}{\partial p_1} &= -\frac{\mu\textsl{\textrm{a}}}{r}
    \biggl[\frac{2\sigma_1}{\mu\beta}{\dot{\bf r}}+\Bigl(\frac{p_2}{r\textsl{\textrm{w}}}-\frac{\sigma_2}{c}
    \Bigr){\bf e}_r+\frac{p_1}{h}{\bf e}_f\biggr]+\frac{1}{r}{\bf s}_1+f_1{\bf e}_f,\\[1ex] 
    \frac{\partial{\dot{\bf r}}}{\partial p_2} &= -\frac{\mu\textsl{\textrm{a}}}{r}
    \biggl[\frac{2\sigma_2}{\mu\beta}{\dot{\bf r}}-\Bigl(\frac{p_1}{r\textsl{\textrm{w}}}-\frac{\sigma_1}{c}
    \Bigr){\bf e}_r+\frac{p_2}{h}{\bf e}_f\biggr]+\frac{1}{r}{\bf s}_2+f_2{\bf e}_f,\\[1ex]
    \frac{\partial{\dot{\bf r}}}{\partial \mathcal{L}} &= \frac{1}{r\nu}\biggl[
    -2\dot{r}\dot{\bf r}+\frac{\mu}{r}\Bigl(1-\frac{r}{\textsl{\textrm{a}}}
    \Bigr){\bf e}_r\biggr]+\frac{1}{r}{\bf s}_3+f_3{\bf e}_f,\\[1ex]
    \frac{\partial{\dot{\bf r}}}{\partial q_1} &= \frac{2}{\gamma}
    (q_2\dot{\bf r}\times{\bf e}_h-\dot{X}{\bf e}_h)+
    f_4{\bf e}_f,\\[1ex]
    \frac{\partial{\dot{\bf r}}}{\partial q_2} &= -\frac{2}{\gamma}
    (q_1\dot{\bf r}\times{\bf e}_h-\dot{Y}{\bf e}_h)+
    f_5{\bf e}_f,
  \end{split}
  \label{eq:ddr_dp}
\end{equation}
where the variable $\textsl{\textrm{w}}$ is defined in ({\ref{eq:w}),
\begin{align*}
    f_0 &= \frac{r}{h}\biggl(\frac{2}{3\nu}\mathscr{U}-\frac{\partial{\mathscr{U}}}{\partial \nu}\biggr),\\[1ex]
    f_i &= -\frac{1}{h}\biggl(\frac{2\sigma_i\textsl{\textrm{a}}}{\beta}\mathscr{U}
    +r\frac{\partial\mathscr{U}}{\partial p_i}\biggr),\quad i=1,2,\\[1ex]
    f_3 &= -\frac{1}{h}\biggl(\frac{2\dot{r}}{\nu}\mathscr{U}+r\frac{\partial\mathscr{U}}
    {\partial \mathcal{L}}\biggr),\\[1ex]
    f_{i+3} &= -\frac{r}{h}\frac{\partial\mathscr{U}}{\partial q_i},\quad i=1,2,        
\end{align*}
and
\begin{align*}
  \sigma_1 &= \frac{1}{r}[\alpha p_1(\varrho-r)-Y],\\[1ex]
  \sigma_2 &= \frac{1}{r}[\alpha p_2(\varrho-r)-X],\\[1ex]
  {\bf s}_i &= \dot{r}\frac{\partial{\bf r}}{\partial p_i}+\frac{h}{r}{\bf e}_h\times
  \frac{\partial{\bf r}}{\partial p_i},\qquad i=1,2,\\[1ex]
  {\bf s}_3 &= \dot{r}\frac{\partial{\bf r}}{\partial\mathcal{L}}+\frac{h}{r}{\bf e}_h\times
  \frac{\partial{\bf r}}{\partial\mathcal{L}}.
\end{align*}
In (\ref{eq:ddr_dp}) the terms $f_i,\,(i=1,\dots,5)$ are equal to zero
if $\mathscr{U}=0$.

If the constant time element $\mathcal{L}_0$ is used instead of $\mathcal{L}$, we have
\begin{align*}
  \frac{\partial{\bf r}}{\partial \nu} &= \frac{1}{\nu}\Bigl(t{\bm\upsilon}-
    \frac{2}{3}{\bf r}\Bigr),\\[1ex]
  \frac{\partial{\bf r}}{\partial \mathcal{L}_0} &= \frac{\partial{\bf r}}{\partial \mathcal{L}},
\end{align*}
and
\begin{align*}
  \frac{\partial{\dot{\bf r}}}{\partial \nu} &= \frac{1}{3\nu}\dot{\bf r}+f_0{\bf e}_f+
  \frac{t}{r^2\nu}\left(\frac{c}{h}(c-h)\,\dot{\bf r}\times{\bf e}_h-\mu{\bf e}_r\right),\\[1ex]
  \frac{\partial{\dot{\bf r}}}{\partial \mathcal{L}_0} &= \frac{\partial{\dot{\bf r}}}{\partial \mathcal{L}}.
\end{align*}
On the other hand, the partial derivatives of ${\bf r}$, $\dot{\bf r}$
with respect to $p_1$, $p_2$, $q_1$, $q_2$ remain the same as in
(\ref{eq:dr_dp}), (\ref{eq:ddr_dp}).

\emph{Remark 3.} The partial derivative of $\mathscr{U}$ with
respect to any element $\chi$ of our set of GEqOE is computed by the
chain rule
\[
\frac{\partial{\mathscr{U}}}{\partial\chi} = \frac{\partial{\mathscr{U}}}{\partial{\bf r}}
\frac{\partial{\bf r}}{\partial\chi}.
\]

\begin{figure*}
  \centering
  \includegraphics[width=0.95\textwidth]{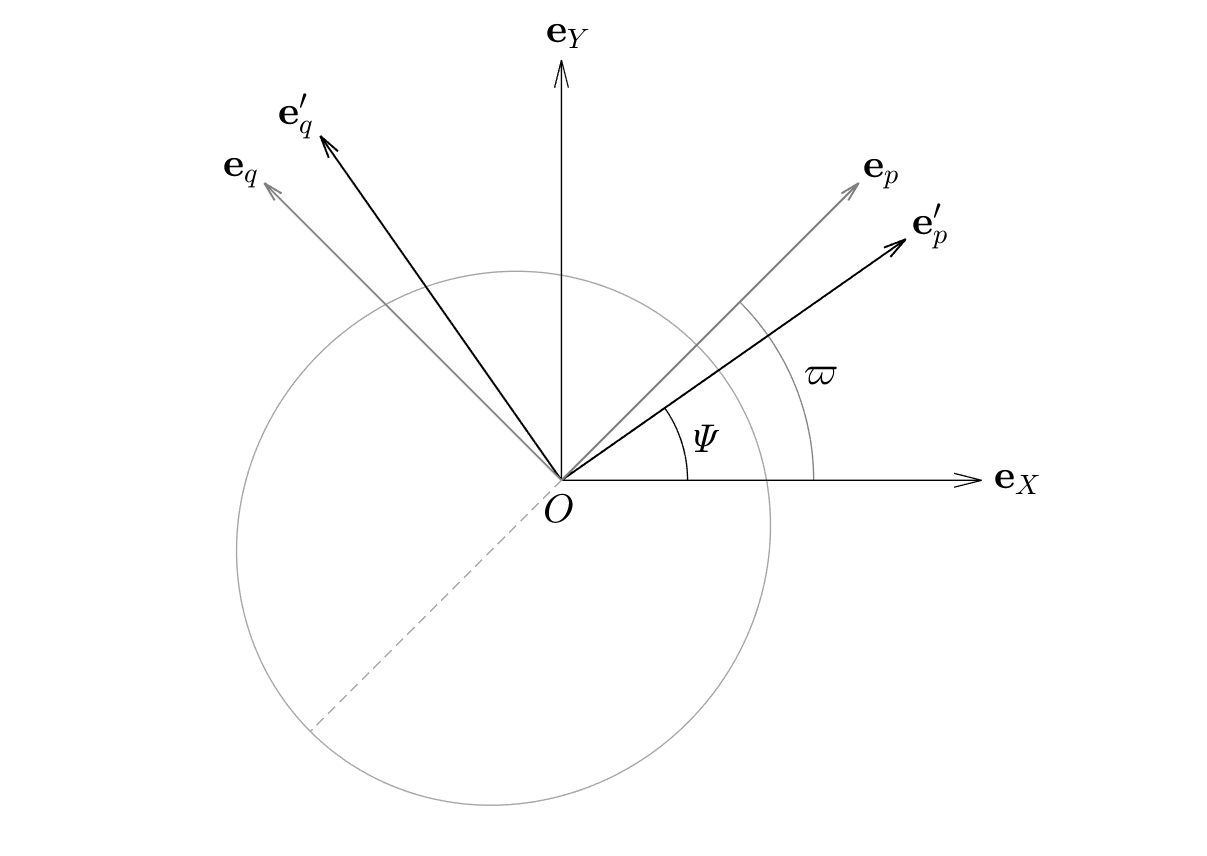}
  \caption{View from the angular momentum vector of the unit vectors
    ${\bf e}_X$, ${\bf e}_Y$ of the equinoctial basis, and of ${\bf
      e}_p'$, ${\bf e}_q'$, which are the generalized counterparts of
    the unit vectors ${\bf e}_p$, ${\bf e}_q$ of the perifocal basis.}
  \label{fig:perifocal}
\end{figure*}

\subsection{Partial derivatives of the GEqOE with respect to position and velocity}
\noindent The inverse of the fundamental matrix for the equinoctial
elements is obtained in \citet{BC_1972} using the Poisson brackets and
the fundamental matrix \citep[see also][]{B_1970}. Here, we proceed as
follows. As concerns the elements $q_1$, $q_2$, we simply put the
expressions given in \citet[][Table III]{BC_1972} for the derivatives
of $p$, $q$ in a suitable form to avoid singularities for small
eccentricities and inclinations \citep[as in][]{SST_1995}. For the
generalized mean motion $\nu$ we started from equation (\ref{eq:enne})
and used the definition of the total energy. The computation for the
elements $p_1$, $p_2$ was done considering equations (\ref{eq:p1_L}),
(\ref{eq:p2_L}) and taking into account (\ref{eq:tl}). Finally,
concerning $\mathcal{L}$ and $\mathcal{L}_0$, equations
(\ref{eq:elle_2}), (\ref{eq:r_2}), (\ref{eq:rrd_2}) were used.

The partial derivatives with respect to the position and velocity
read\footnote{We found a typo in the expression of
  $\partial\lambda_0/\partial{\bf x}$ reported in \citet[][Table
    III]{BC_1972}: $\alpha_5/\alpha_6$ has to be replaced by
  $\alpha_5/\alpha_4$.}
\begin{equation}
  \begin{split}
    \frac{\partial \nu}{\partial{\bf r}} &=
    -\frac{3\textsl{\textrm{a}}\nu}{r^2}{\bf e}_r^T+{\bf f}_0,\\[1ex]
    \frac{\partial p_1}{\partial{\bf r}} &= \frac{1}{r}(p_1+\sin L){\bf e}_r^T-\frac{h}{cr}
    \biggl[\Bigl(2-\frac{c}{h}\Bigr)p_2+\frac{X}{\textsl{\textrm{a}}}\biggr]{\bf e}_f^T
    -\frac{p_2\Lambda}{h}{\bf e}_h^T+{\bf f}_1,\\[1ex]
    \frac{\partial p_2}{\partial{\bf r}} &= \frac{1}{r}(p_2+\cos L){\bf e}_r^T+\frac{h}{cr}
    \biggl[\Bigl(2-\frac{c}{h}\Bigr)p_1+\frac{Y}{\textsl{\textrm{a}}}\biggr]{\bf e}_f^T
    +\frac{p_1\Lambda}{h}{\bf e}_h^T+{\bf f}_2,\\[1ex]
    \frac{\partial \mathcal{L}}{\partial{\bf r}} &= \frac{\dot{r}}{cr}(\varrho\alpha-r \beta)
    {\bf e}_r^T-\frac{h}{cr}\Bigl(2-\frac{c}{h}+\frac{\alpha}{\textsl{\textrm{a}}}(\varrho-r)\Bigr){\bf e}_f^T
    -\frac{\Lambda}{h}{\bf e}_h^T+{\bf f}_3,\\[1ex]
    \frac{\partial q_1}{\partial{\bf r}} &=
    -\frac{\gamma\dot{Y}}{2h}{\bf e}_h^T,\\[1ex]
    \frac{\partial q_2}{\partial{\bf r}} &=
    -\frac{\gamma\dot{X}}{2h}{\bf e}_h^T,
  \end{split}
  \label{eq:dp_dr}
\end{equation}
and
\begin{equation}
  \begin{split}
    \frac{\partial \nu}{\partial{\dot{\bf r}}} &=
    -\frac{3}{\sqrt{\mu\textsl{\textrm{a}}}}\dot{\bf r}^T,\\[1ex]
    \frac{\partial p_1}{\partial{\dot{\bf r}}} &=
    -\frac{c}{\mu}\cos L\,{\bf e}_r^T+\frac{h}{\mu}\Bigl(2\sin L
    -\frac{\dot{r}}{c}X\Bigr){\bf e}_f^T+\frac{p_2\lambda}{h}{\bf e}_h^T,\\[1ex]
    \frac{\partial p_2}{\partial{\dot{\bf r}}} &=
    \frac{c}{\mu}\sin L\,{\bf e}_r^T+\frac{h}{\mu}\Bigl(2\cos L
    +\frac{\dot{r}}{c}Y\Bigr){\bf e}_f^T-\frac{p_1\lambda}{h}{\bf e}_h^T,\\[1ex]
    \frac{\partial \mathcal{L}}{\partial{\dot{\bf r}}} &= \Bigl(\frac{c}{\mu r}\alpha(r-\varrho)
    -\frac{2r}{\sqrt{\mu\textsl{\textrm{a}}}}\Bigr){\bf e}_r^T+\frac{h\dot{r}}{c\mu}\alpha
    (\varrho+r){\bf e}_f^T+\frac{\lambda}{h}{\bf e}_h^T,\\[1ex]
    \frac{\partial q_1}{\partial{\dot{\bf r}}} &=
    \frac{\gamma Y}{2h}{\bf e}_h^T,\\[1ex]
    \frac{\partial q_2}{\partial{\dot{\bf r}}} &=
    \frac{\gamma X}{2h}{\bf e}_h^T,
  \end{split}
  \label{eq:dp_ddr}
\end{equation}
where
\begin{align*}
  {\bf f}_0 &= -\frac{3}{\sqrt{\mu\textsl{\textrm{a}}}}\frac{\partial{\mathscr{U}}}{\partial{\bf r}},
  \\[1ex]
  {\bf f}_1 &= \frac{1}{\mu\varrho}\left[(\varrho+r)Y+p_1r^2\right]
  \frac{\partial\mathscr{U}}{\partial{\bf r}},\\[1ex]
  {\bf f}_2 &= \frac{1}{\mu\varrho}\left[(\varrho+r)X+p_2r^2\right]
  \frac{\partial\mathscr{U}}{\partial{\bf r}},\\[1ex]
  {\bf f}_3 &= \frac{r\dot{r}}{c\mu}(\varrho+r)\alpha\frac{\partial\mathscr{U}}{\partial{\bf r}},
\end{align*}
and
\[
\lambda = Yq_2-Xq_1,\qquad \Lambda = \dot{Y}q_2-\dot{X}q_1.
\]
The vectors ${\bf f}_i$ ($i=0,\ldots,3$) are equal to zero if
$\mathscr{U}=0$.

Finally, the partial derivatives of the constant time element
$\mathcal{L}_0$ are given by
\begin{align*}
  \frac{\partial \mathcal{L}_0}{\partial{\bf r}} &= \frac{\partial \mathcal{L}}{\partial{\bf r}}
  -3t\Bigl(\frac{\alpha}{r^2}{\bf e}_r^T-\frac{1}{\sqrt{\mu\textsl{\textrm{a}}}}\frac{\partial\mathscr{U}}
  {\partial{\bf r}}\Bigr),\\[1ex]
  \frac{\partial \mathcal{L}_0}{\partial{\dot{\bf r}}} &= \frac{\partial \mathcal{L}}{\partial{\dot{\bf r}}}
  +\frac{3t}{\sqrt{\mu\textsl{\textrm{a}}}}\dot{\bf r}^T.
\end{align*}

\emph{Remark 4.} If $\mathcal{L}_0$ is employed in place of
$\mathcal{L}$, terms that are linear in time are introduced in the
expressions of $\partial{{\bf r}}/\partial{\nu}$, $\partial{\dot{\bf
    r}}/\partial{\nu}$, $\partial{\mathcal{L}_0}/\partial{{\bf r}}$,
$\partial{\mathcal{L}_0}/\partial{\dot{\bf r}}$.

\section{Numerical tests}
\label{sec:results}
We employ the generalized equinoctial elements (GEqOE) $\nu$, $p_1$,
$p_2$, $\mathcal{L}$, $q_1$, $q_2$ (Section \ref{sec:summary}) to
propagate the motion of an artificial satellite around the
Earth. Starting from the case in which only the perturbation due to
the $J_2$ zonal harmonic of the geopotential is considered, we then
add the third-body gravitational attraction of the Moon and Sun. The
GEqOE are compared to the alternate equinoctial orbital elements
(AEqOE) presented in \citet[][Section 10.4]{HAP_2011} and Cowell's
method. We deliberately select a very simple numerical integrator,
which is a fourth order Runge-Kutta with a fixed step size, in order
to highlight the impact of the particular set of elements on the
propagation performance. The errors are computed by taking as
\emph{true} the solution obtained by using the Dromo(PC) formulation
presented in \citet{BauB_2014} and an adaptive step size Runge-Kutta
Dormand-Prince 5(4)7FM method described in \cite{DormandePrince_1980}
with a tolerance of $10^{-13}$.

\subsection{The main problem}
\label{sec:mainp}
Let us introduce an Earth-centered inertial reference frame. In
particular, we denote by $Oz$ the axis pointing to the North Pole and
by ${\bf e}_z$ the corresponding unit vector. We consider here the
unrealistic case in which the perturbing force ${\bf F}$ (see equation
\ref{eq:F}) is given by
\[
{\bf F} = \nabla\mathscr{R}({\bf r}),
\]
where $\mathscr{R}$ is the disturbing potential associated with the
$J_2$ term of the Earth's gravitational field: 
\begin{equation}
  \mathscr{R}=-\mathscr{U}=\frac{A}{r^3}(1-3\hat{z}^2),
  \label{eq:R}
\end{equation}
with
\[
A=\frac{GMJ_2r_{eq}^2}{2},\qquad\hat{z}=\frac{z}{r},
\]
and $z={\bf r}\cdot{\bf e}_z$. The quantity $r_{eq}$ denotes the
equatorial radius of the Earth. We have
\[
{\bf F} = -\frac{3A}{r^4}\bigl[(1-5\hat{z}^2){\bf e}_r+2\hat{z}{\bf e}_z\bigr].
\]
In the differential equations of the generalized equinoctial elements
the perturbing force appears only as $2\mathscr{U}-rF_r$,
$r\hat{w}_hF_h$, $rF_h$, where $F_r={\bf F}\cdot{\bf e}_r$,
$F_h={\bf F}\cdot{\bf e}_h$. A straightforward computation yields
\begin{align*}
  & 2\mathscr{U}-rF_r = -\mathscr{U},\\
  & r\hat{w}_hF_h = \frac{3A}{r^3}\hat{z}^2(1-q_1^2-q_2^2) =
  \frac{3A}{hr^3}\hat{z}^2\,\mathscr{C}(1+q_1^2+q_2^2),\\
  & rF_h = -\frac{6A}{hr^3}\hat{z}\,\mathscr{C},
\end{align*}
where we used the relation
\[
\cos i = \frac{\mathscr{C}}{h},
\]
and $\mathscr{C}$ is the component of the angular momentum vector
along $Oz$, which is a first integral of the motion.
\noindent Taking into account that the total energy $\mathscr{E}$ is
also a first integral in this case, the time derivatives of
the GEqOE become
\begin{align*}
  \dot{\nu} & = 0,\\[1ex]
  \dot{p}_1 & = p_2\biggl(\frac{h-c}{r^2}-I\hat{z}\biggr)-\frac{1}{c}
  \Bigl(\frac{r\dot{r}}{c}p_1+\tilde{\varsigma}p_2+\varsigma\cos L\Bigr)\mathscr{U},\\[1ex]
  \dot{p}_2 & = -p_1\biggl(\frac{h-c}{r^2}-I\hat{z}\biggr)-\frac{1}{c}
  \Bigl(\frac{r\dot{r}}{c}p_2-\tilde{\varsigma}p_1-\varsigma\sin L\Bigr)\mathscr{U},\\[1ex]
  \dot{\mathcal{L}}_0 & = \frac{h-c}{r^2}-I\hat{z}-\frac{1}{c}
  \biggl[\frac{1}{\alpha}+\alpha\Bigl(1-\frac{r}{\textsl{\textrm{a}}}\Bigr)\biggr]\mathscr{U},\\[1ex]
  \dot{q}_1 & = -I\sin L,\\[1ex]
  \dot{q}_2 & = -I\cos L,
\end{align*}
where
\[
I = \frac{3A}{h^2r^3}\hat{z}\,\mathscr{C}(1+q_1^2+q_2^2).
\]
The meaning of the variables $\textsl{\textrm{a}}$, $\alpha$,
$\tilde{\varsigma}$, $\varsigma$ is explained in (\ref{eq:ga}),
(\ref{eq:alpha}), (\ref{eq:aux}). In the equations above the
quantities $\mathscr{C}$, $\textsl{\textrm{a}}$, $\nu$ remain constant
along the motion and their values are determined by the initial
position (${\bf r}$) and velocity ($\dot{\bf r}$). In particular, the
value of the element $\nu$ is computed by (see equation \ref{eq:enne})
\[
\nu=\frac{1}{\mu}\biggl[\frac{2A}{r_*^3}(1-3\hat{z}_*^2)-|{\dot{\bf r}}_*|^2\biggr]^{3/2},
\]
where $r_*$, $\hat{z}_*$, ${\dot{\bf r}}_*$ refer to the starting epoch of the propagation.

\begin{table}
  \centering
  \caption{Initial conditions for the numerical tests: $r_p$ is the
    radius of perigee in km, $e$, $i$, $\Omega$, $\omega$, $M_0$
    are Keplerian elements. Angular variables are given in degrees and
    expressed with respect the J2000 frame.}
  \label{tab:ic}
  \begin{tabular}{llllrrl}
    \hline\noalign{\smallskip}
    & $r_p$ & $e$ & $i$ & $\Omega$ & $\omega$ & $M_0$\\
    \noalign{\smallskip}\hline\noalign{\smallskip}
    1) &7178.1366  &0  &45  &0  &0  &0 \\[1ex]
    2) &7178.1366 &0  &90  &0  &0  &0 \\[1ex]
    3) &26000  &0.74  &63.4  &30  &270  &0 \\
    \noalign{\smallskip}\hline
  \end{tabular}
\end{table}

The initial conditions specified in the row labeled 1) in Table
\ref{tab:ic}, which correspond to a low Earth orbit, are propagated
for 12 days. Figure \ref{fig:J2_1} shows the position error
at the final time obtained with different values of the integration
step size. The same initial conditions are then propagated for 365
days using a step size of 1 minute. The time evolution of the errors
in the position and total energy are displayed in Figure
\ref{fig:J2_23}. It is evident that the GEqOE are much better than the
EqOE and Cowell's method for this test case.

\begin{figure*}
  \centering
  \includegraphics[width=0.7\textwidth]{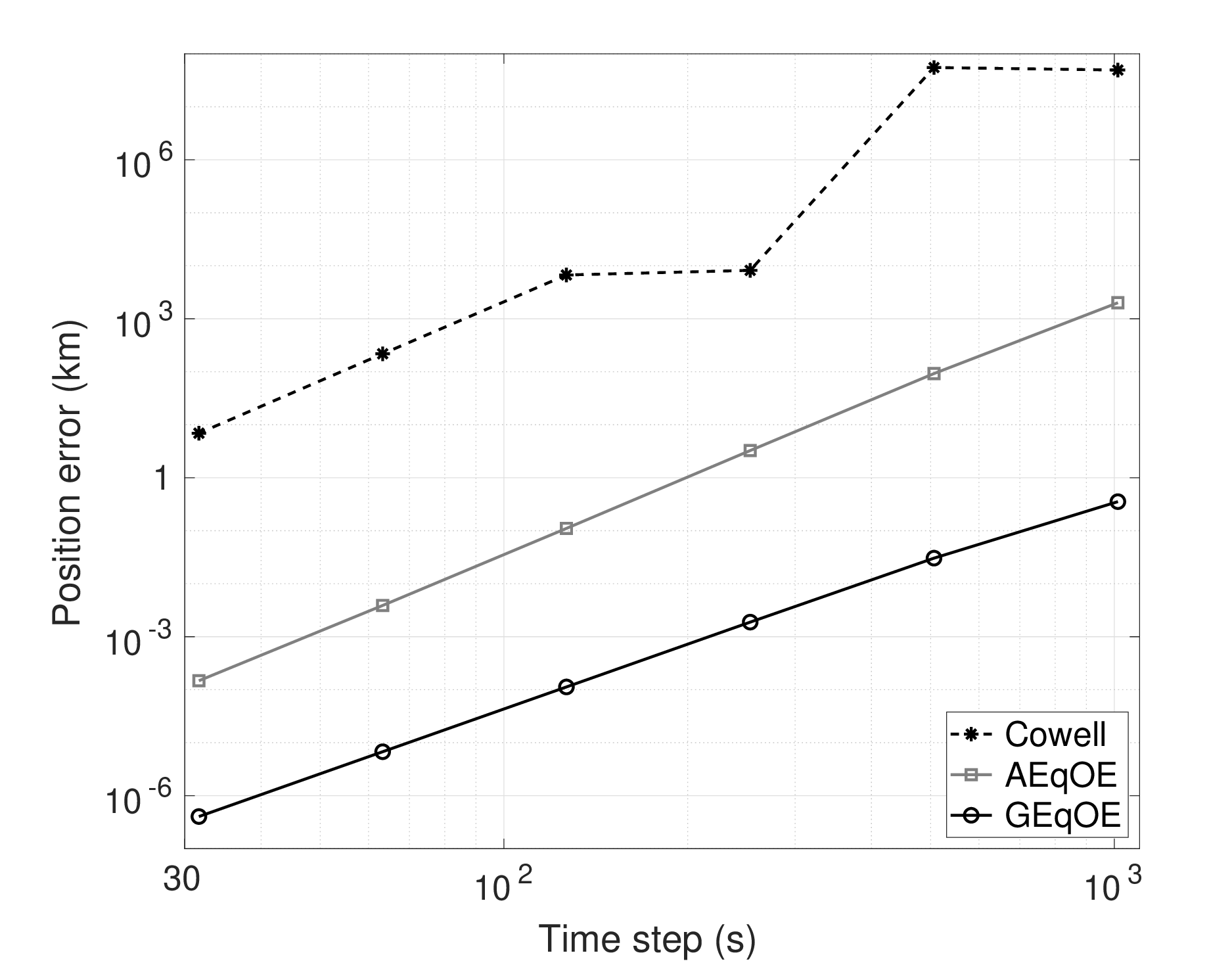}
  \caption{Position error after 12 days of propagation for increasing
    values of the integration step size. The initial conditions
    correspond to the low Earth orbit reported in the row 1) of Table
    \ref{tab:ic}. Only the perturbation arising from the $J_2$ zonal
    harmonic of the geopotential is considered. Note that a
    logarithmic scale is applied to both axes.}
  \label{fig:J2_1}
\end{figure*}

\begin{figure*}
  \centering
  \includegraphics[width=0.7\textwidth]{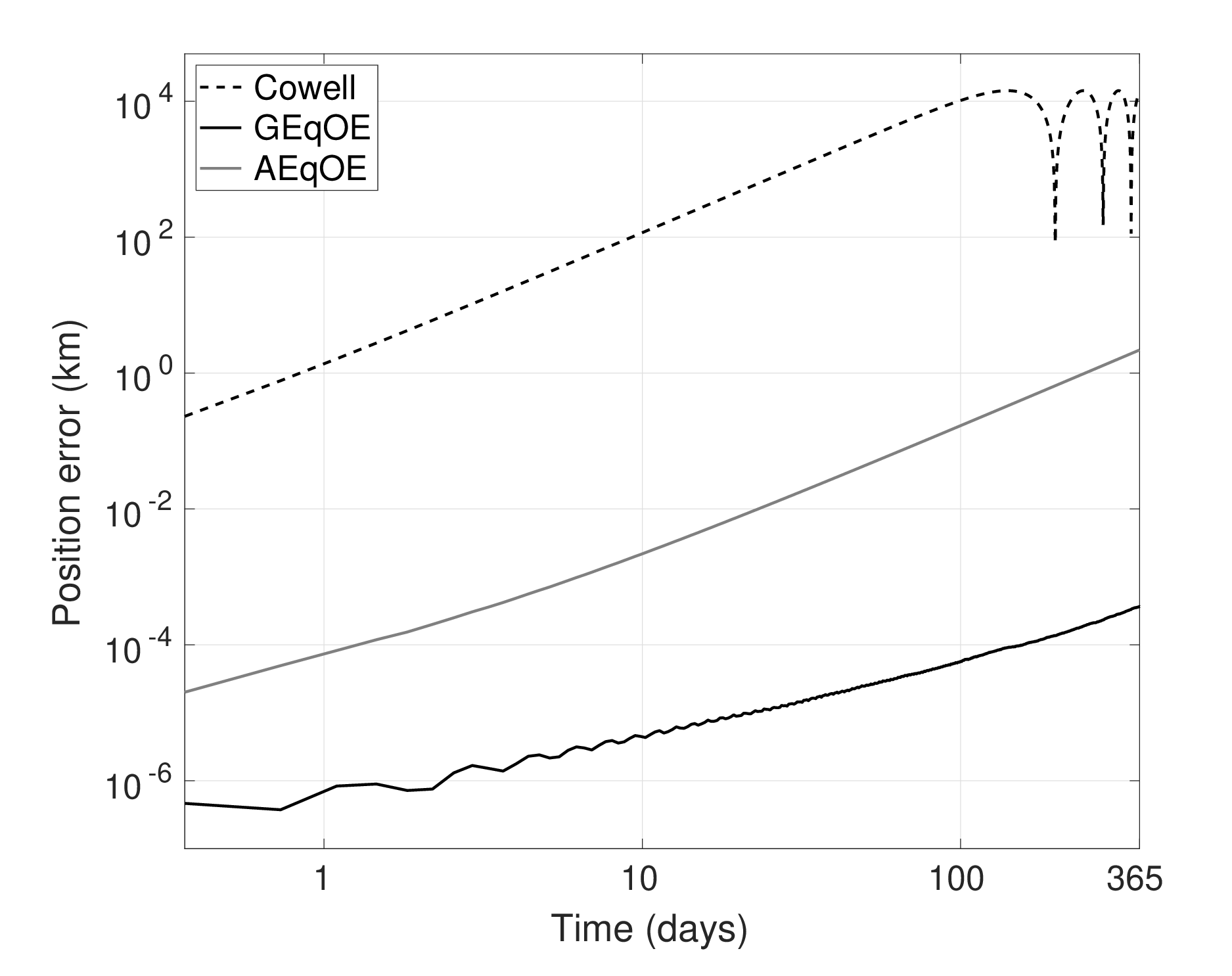}
  \includegraphics[width=0.7\textwidth]{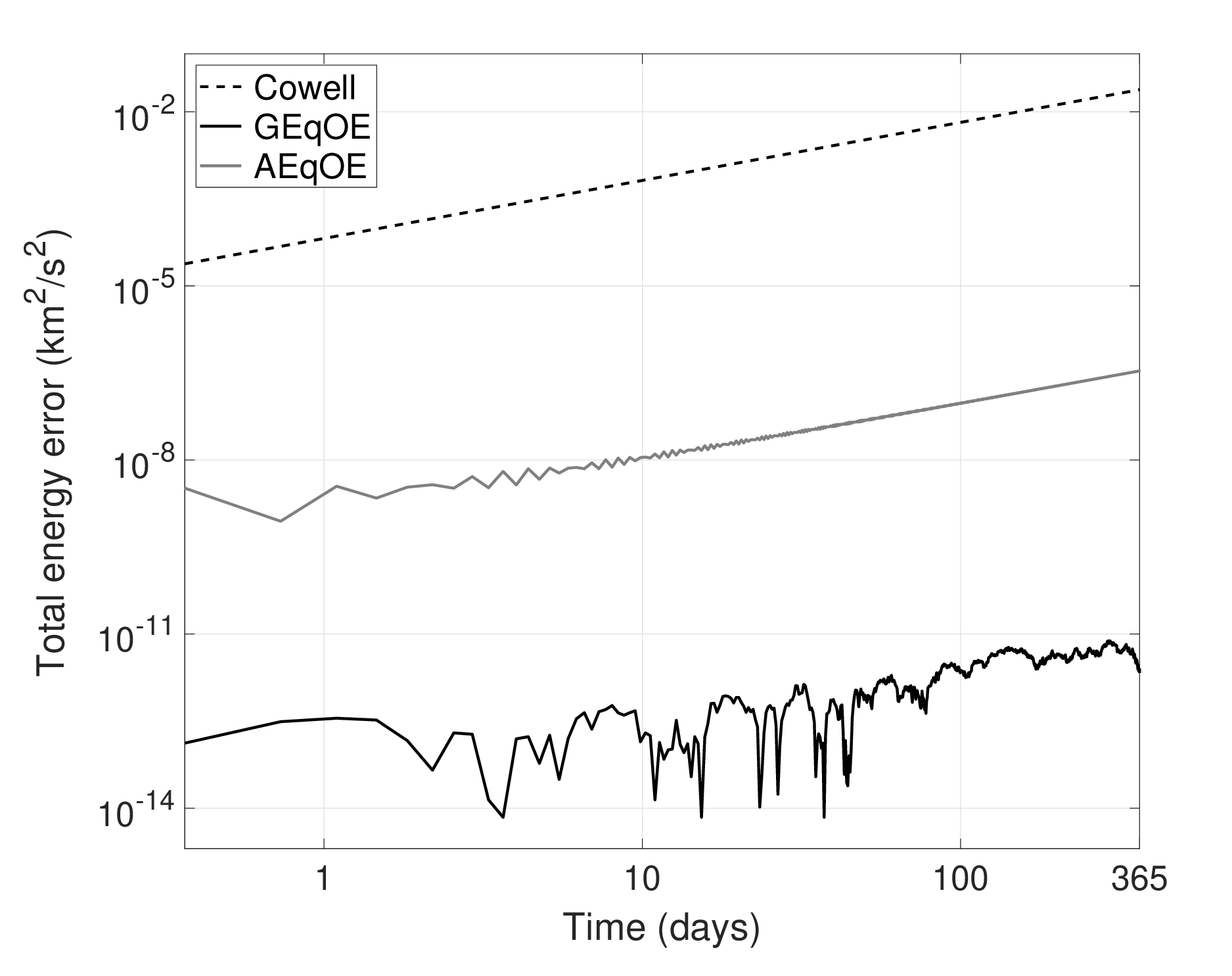}
  \caption{Time evolution of (\emph{top}) the position error and
    (\emph{bottom}) the error of the total energy throughout 365 days
    of propagation of the initial conditions reported in the row 1) of
    Table \ref{tab:ic}. Only the perturbation arising from the $J_2$
    zonal harmonic of the geopotential is considered. Note that a
    logarithmic scale is applied to both axes.}
  \label{fig:J2_23}
\end{figure*}

\subsection{$J_2$ and third-body perturbations}
Third-body perturbing forces due to the gravitational attraction of
the Moon and Sun are switched on in the following tests. Since we
decided to not derive them from a potential, these forces determine
the vector ${\bf P}$ in equation (\ref{eq:F}). On the other hand, the
force exerted by the Earth's oblateness is derived from the potential
$\mathscr{R}$ introduced in (\ref{eq:R}).

The same performance plots shown in Section \ref{sec:mainp} are
obtained for the initial conditions reported in the rows 2) and 3) of
Table \ref{tab:ic}, which correspond to a low Earth orbit and a
Molniya orbit. The initial epoch of the propagations is January 1,
2020 UTC. For a short-term propagation (in the order of a few days),
we display in Figure \ref{fig:J2_MS_12} the variation of the position
error referred to the final time as the step size of the integrator is
enlarged. Then, for a long-term propagation (in the order of hundreds
of days) Figure \ref{fig:J2_MS_34} shows the time evolution of the
position error using a step size of 1 minute. The timespan of the
Molniya orbit propagation was chosen about 7 times larger in order to
have the same number of revolutions as the case of the low Earth
orbit.

From these results we see that third-body perturbations do not
deteriorate the performance of the GEqOE, when compared to the case in
which only a conservative perturbation is present. The generalized
equinoctial orbital elements brings a substantial improvement with
respect to the AEqOE and Cowell's method.

Because the Molniya orbit is quite eccentric, it is reasonable to use
also a variable step size numerical integrator: we chose the same
Runge-Kutta Dormand-Prince method that provides the \emph{true}
solution. The relative tolerance controls the selection of the step
size and tighter tolerances imply shorter steps. The total number of
evaluations of the vector field at the end of the propagation can be
taken as a measure of the computational cost. In Figure
\ref{fig:J2_MS_5} we show for decreasing values of the relative
tolerance the number of evaluations and the corresponding maximum
position error reached in a 85.6 days propagation interval. We denote
with GEqOE(c) the set of generalized orbital elements in which
$\mathcal{L}_0$ (see Section \ref{sec:cte}) is employed instead of
$\mathcal{L}$. While in the previous numerical tests GEqOE and
GEqOE(c) exhibit an almost identical performance, in this test the
latter formulation is better: it is faster and decreases the minimum
achievable position error. We see that the formulations GEqOE and
GEqOE(c) are much more efficient than AEqOE and Cowell's method.

\begin{figure*}
  \centering
  \includegraphics[width=0.7\textwidth]{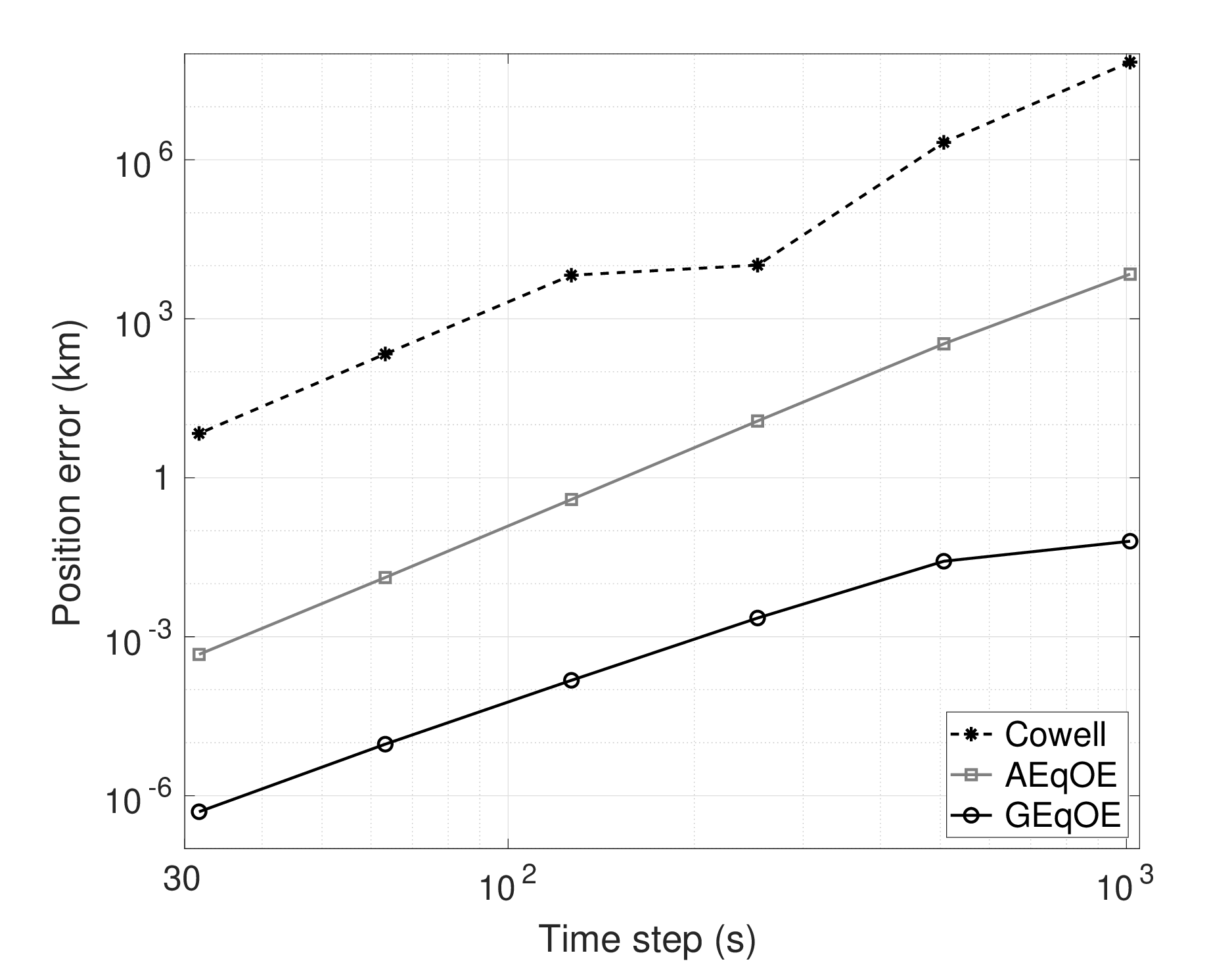}
  \includegraphics[width=0.7\textwidth]{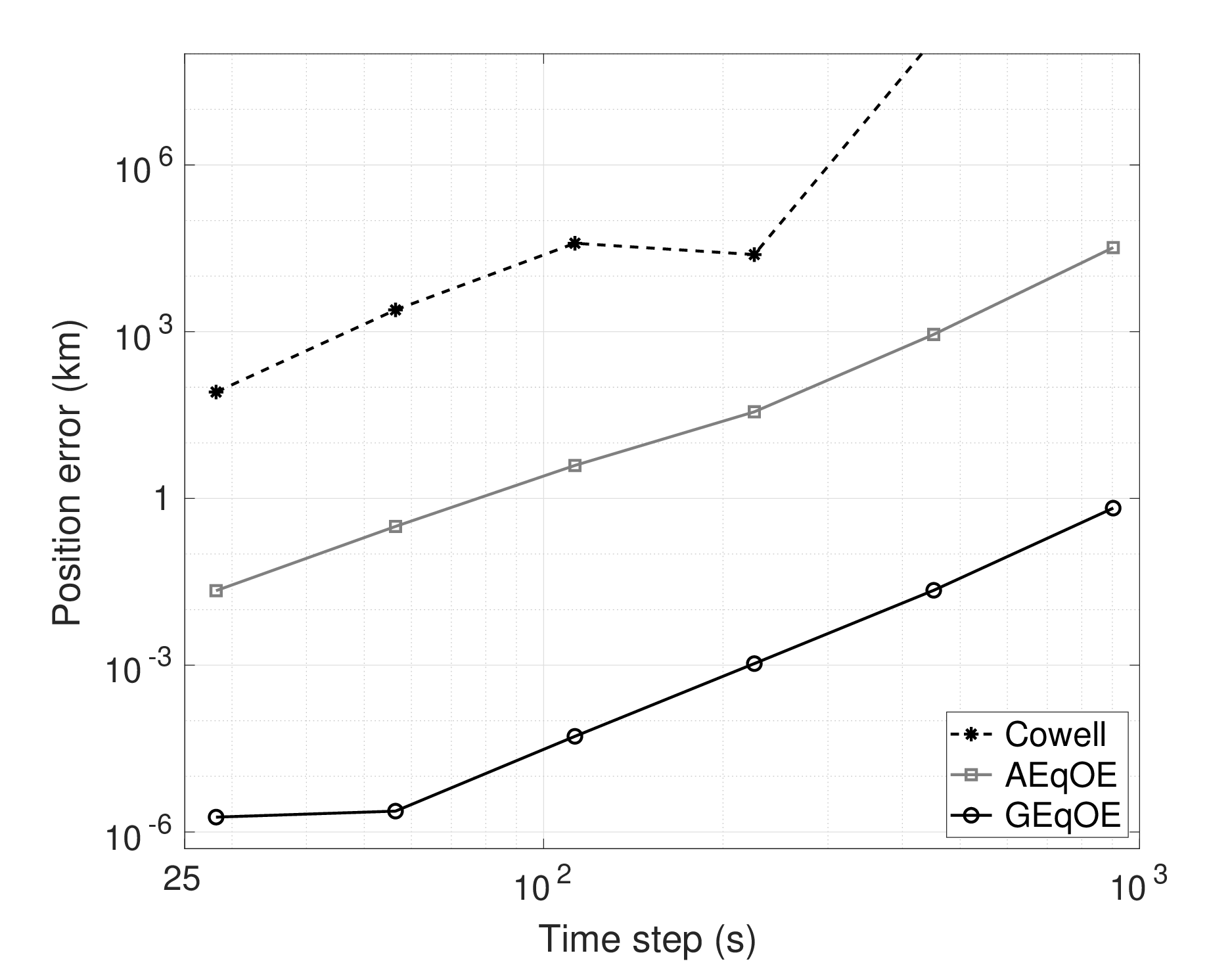}
  \caption{Position error for increasing values of the integration
    step size. The initial conditions correspond to (\emph{top}) the
    low Earth orbit and (\emph{bottom}) the Molniya orbit reported in
    the rows 2) and 3), respectively, of Table \ref{tab:ic}. The
    former is propagated for 12 days, the latter for 85.6
    days. Perturbations due to the $J_2$ zonal harmonic of the
    geopotential and the attraction of the Moon and Sun are
    considered. Note that a logarithmic scale is applied to both
    axes.}
  \label{fig:J2_MS_12}
\end{figure*}

\begin{figure*}
  \centering
  \includegraphics[width=0.7\textwidth]{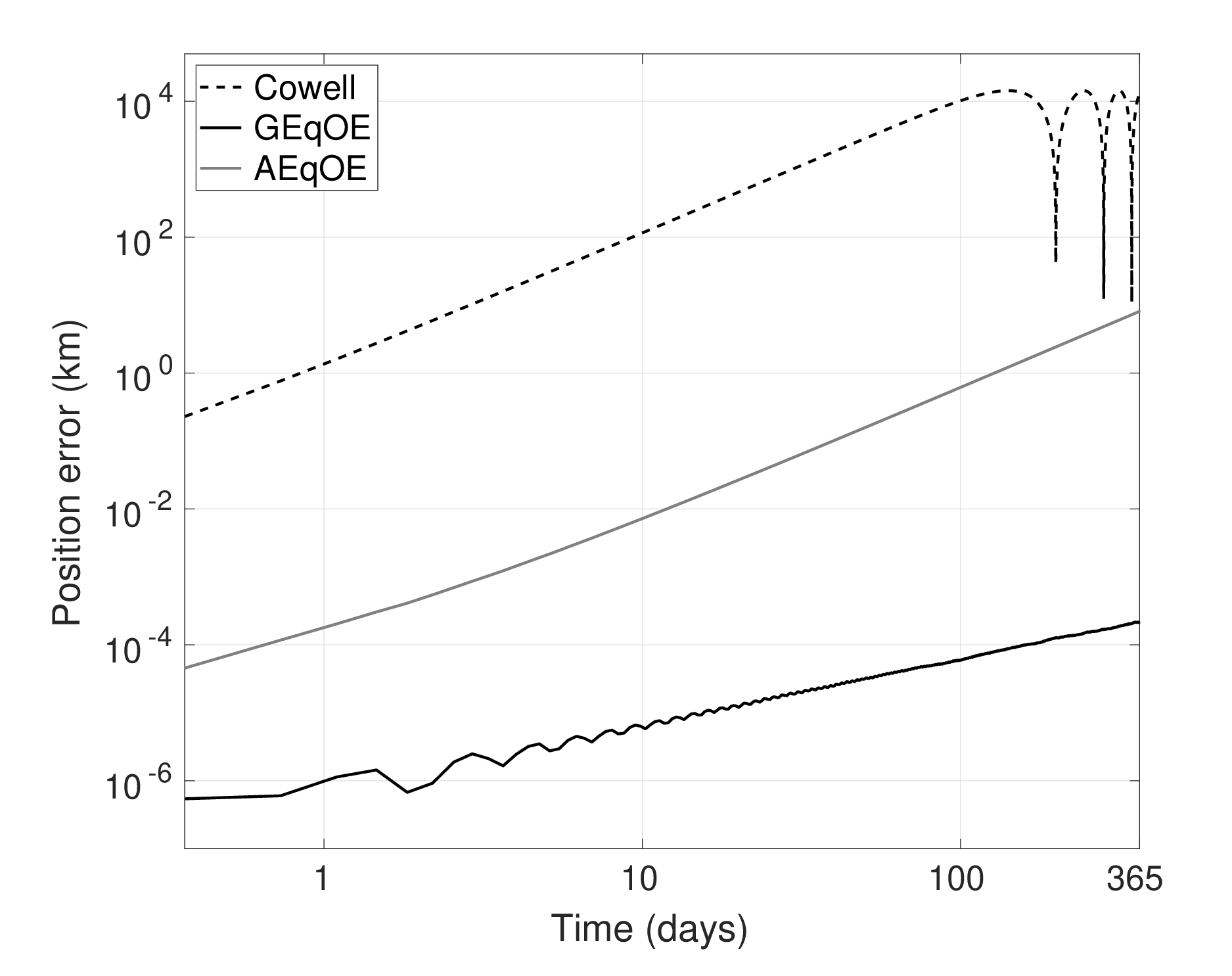}
  \includegraphics[width=0.7\textwidth]{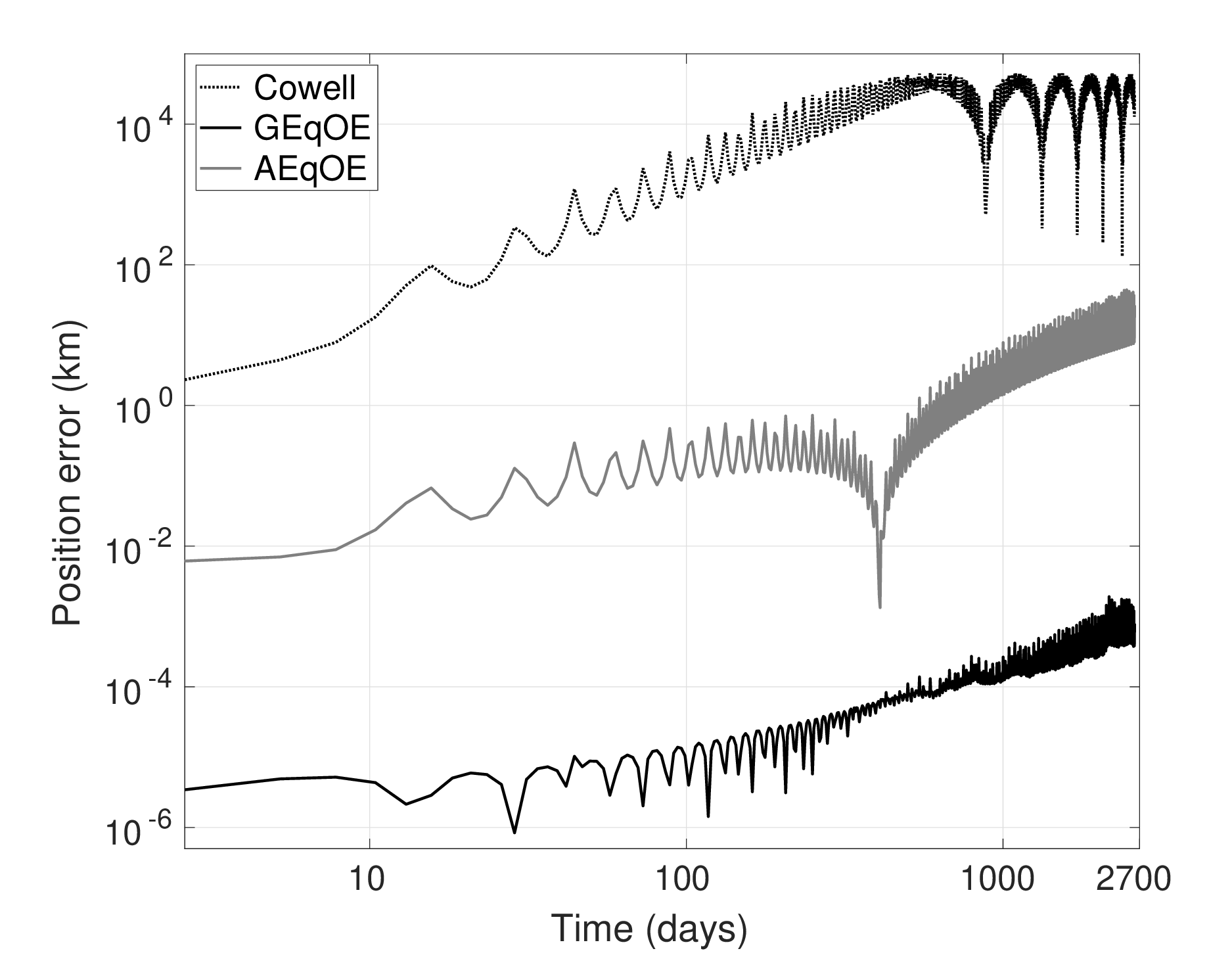}
  \caption{Time evolution of the position error. The initial
    conditions correspond to (\emph{top}) the low Earth orbit and
    (\emph{bottom}) the Molniya orbit reported in the rows 2) and 3),
    respectively, of Table \ref{tab:ic}. The former is propagated for
    365 days, the latter for 2604 days. Perturbations due to the $J_2$
    zonal harmonic of the geopotential and the attraction of the Moon
    and Sun are considered. Note that a logarithmic scale is applied
    to both axes.}
  \label{fig:J2_MS_34}
\end{figure*}

\begin{figure*}
  \centering
  \includegraphics[width=0.7\textwidth]{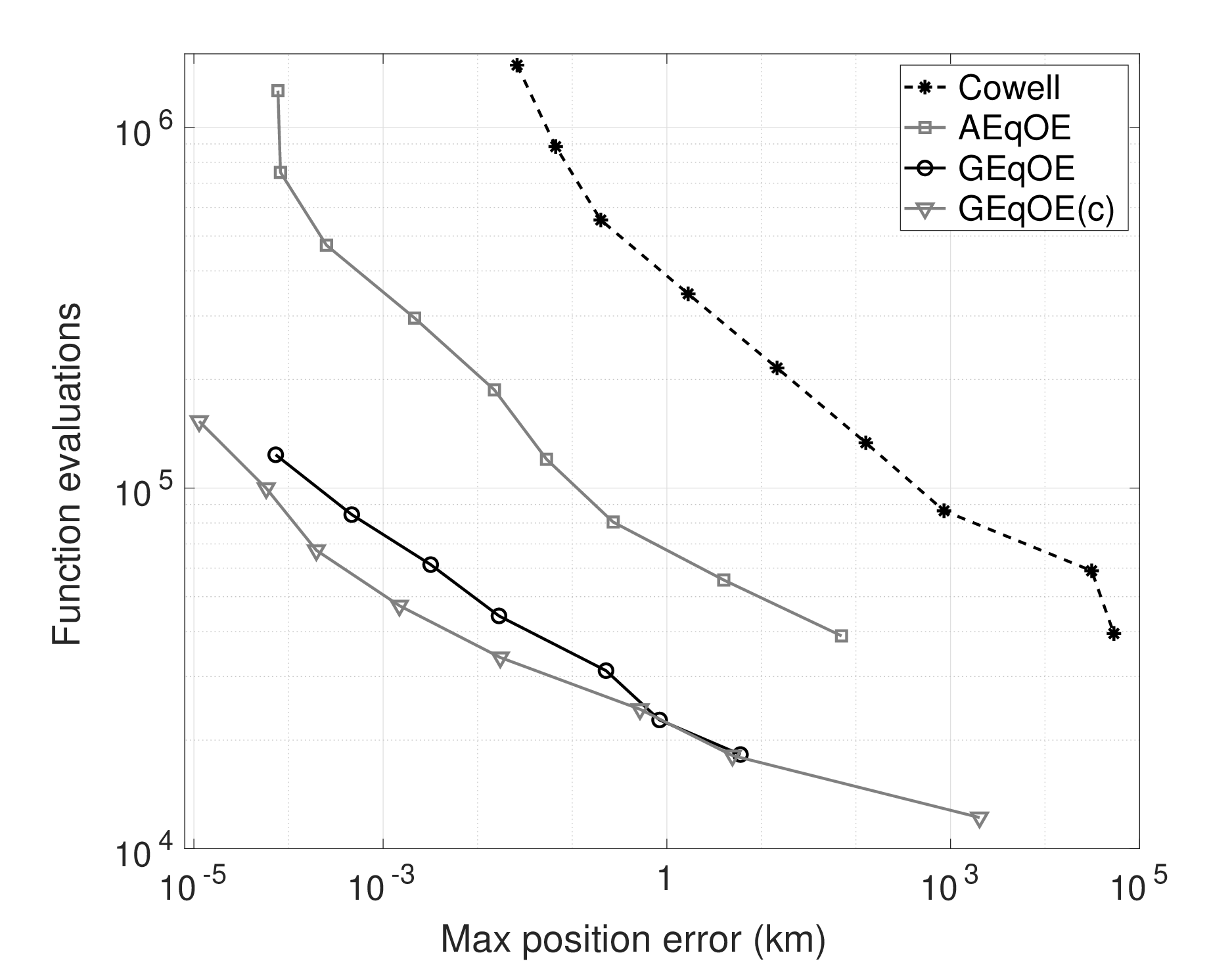}
  \caption{Function evaluations (see the text) versus maximum position
    error in 85.6 days of propagation for decreasing (right to left)
    values of the integrator relative tolerance. The initial
    conditions correspond to the Molniya orbit reported in the row 3)
    of Table \ref{tab:ic}. Perturbations due to the $J_2$ zonal
    harmonic of the geopotential and the attraction of the Moon and
    Sun are considered. Note that a logarithmic scale is applied to
    both axes.}
  \label{fig:J2_MS_5}
\end{figure*}

\section{Conclusions and future work}
We have introduced six orbital elements that generalize the classical
equinoctial elements in the presence of perturbations that are
derivable from a disturbing potential. The latter appears in the
intrinsic definition of the new elements, through the generalization
of different orbital motion quantities. The new elements are defined
for a negative value of the total energy, and a positive value of the
effective potential energy. They are non-singular for circular and
equatorial trajectories, and are affected by the same singularities of
their classical counterpart (retrograde equatorial and rectilinear
orbits). Equations of motions, transformation from and to Cartesian
coordinates are provided along with the associated Jacobian matrices.

Representative propagation tests for low Earth orbits show a dramatic
increase in performance (accuracy and computational cost) for a
propagator based on the new elements compared to the alternate
equinoctial orbital elements as well as Cartesian coordinates. Ongoing
research is focused on the application of the proposed elements to the
problem of uncertainty propagation.



\section*{Acknowledgements}
  G. Ba\`u acknowledges the projects MIUR-PRIN 20178CJA2B titled ``New
  frontiers of Celestial Mechanics: theory and applications'', and PRA
  2020-82 titled ``Sistemi dinamici in logica, geometria, fisica
  matematica e scienza delle costruzioni''. Moreover, he acknowledges
  the INdAM group ``Gruppo Nazionale per la Fisica Matematica''.  The
  views expressed are those of the authors and do not necessarily
  represent the views of ispace, inc.

\section*{Conflict of Interest}
The authors declare that they have no conflict of interest.

\appendix
\section{Expressions of $\sin\mathcal{K}$, $\cos\mathcal{K}$}
\label{sec:sKcK}
Note that from (\ref{eq:Psi}), (\ref{eq:K}) we can
write
\[
\mathcal{K} = \mathcal{L}-(\theta-G).
\]
Let us consider the well-known relation
\[
\tan\left(\frac{\theta-G}{2}\right) = \frac{\sin\theta-\sin G}{\cos\theta+\cos G}.
\]
After replacing $\sin\theta$, $\cos\theta$, $\sin G$, $\cos G$ with
the expressions that can be derived from (\ref{eq:cth}),
(\ref{eq:sth}), (\ref{eq:r}), (\ref{eq:rrd}), respectively, we find
\[
\tan\left(\frac{\theta-G}{2}\right) = \frac{r\dot{r}}{c+\textsl{\textrm{w}}r},
\]
where $\textsl{\textrm{w}}$ is defined by equation (\ref{eq:w}). Taking
into account the formulae
\[
\sin(2\arctan\iota) = \frac{2\iota}{1+\iota^2},\quad
\cos(2\arctan\iota) = \frac{1-\iota^2}{1+\iota^2},
\]
where in our case
\[
\iota = \frac{r\dot{r}}{c+\textsl{\textrm{w}}r},
\]
we get
\begin{align*}
  \sin\mathcal{K} & = \frac{1}{1+\iota^2}\left[(1-\iota^2)\sin\mathcal{L}-
    2\iota\cos\mathcal{L}\right],\\
  \cos\mathcal{K} & = \frac{1}{1+\iota^2}\left[(1-\iota^2)\cos\mathcal{L}+
    2\iota\sin\mathcal{L}\right].
\end{align*}
Finally, with the aid of the definition of the total energy in terms
of $r$, $\dot{r}$, $c$, $\mathscr{U}$, it is possible to show that
\begin{align*}
  \frac{2\iota}{1+\iota^2} = \dot{r}\frac{c+\textsl{\textrm{w}}r}{\mu + c\textsl{\textrm{w}}},\quad
  \frac{1-\iota^2}{1+\iota^2} = 1-\frac{r\dot{r}^2}{\mu + c\textsl{\textrm{w}}}.
\end{align*}

\section{Time derivatives of $p_1$, $p_2$}
\label{sec:dp12dt}
From equations (\ref{eq:p1}), (\ref{eq:p2}) we have
\begin{align}
  \dot{p}_1 & =\dot{g}\sin\Psi+g\dot{\Psi}\cos\Psi,\label{eq:dp1_2}\\
  \dot{p}_2 & =\dot{g}\cos\Psi-g\dot{\Psi}\sin\Psi.\label{eq:dp2_2}
\end{align}

For the time derivative of $g$ we use relation (\ref{eq:ge}) and
so we need the expressions of $\dot{\mathscr{E}}$, $\dot{c}$. The
former is given in (\ref{eq:dE}), the latter is derived from the
definition of $c$ provided in (\ref{eq:c}) and results:
\begin{equation}
  \dot{c} = \frac{1}{c}[r^2\dot{\mathscr{E}}+r\dot{r}(2\mathscr{U}-rF_r)].
  \label{eq:dc}
\end{equation}
Then, we find
\[
\dot{g} = \frac{1}{\mu^2g}\left[(c^2+2\mathscr{E}r^2)
  \dot{\mathscr{E}}+2\mathscr{E}r\dot{r}(2\mathscr{U}-rF_r)\right].
\]
After writing $2\mathscr{E}$ as a function of $c$, $g$ by means of
relation (\ref{eq:ge}), and using
\[
r = \frac{c^2}{\mu(1+g\cos\theta)},\qquad \dot{r} = \frac{\mu}{c}g\sin{\theta},
\]
which directly follows from (\ref{eq:cth}), (\ref{eq:sth}), we obtain
\begin{equation}
  \dot{g} = \frac{r}{\mu}(\tilde{\varsigma}\cos\theta+\varsigma g)\dot{\mathscr{E}}+
  \frac{g^2-1}{c}\varsigma\sin\theta(2\mathscr{U}-rF_r),\label{eq:dg}
\end{equation}
where $\varsigma$, $\tilde{\varsigma}$ are introduced in (\ref{eq:aux}).

From the definition of $\Psi$ we write
\begin{equation}
  g\dot{\Psi} = g\dot{L} - g\dot{\theta}.
  \label{eq:gdPsi}
\end{equation}
The expression of $\dot{L}$ can be derived, for example,
from \citet[][eqs. 10.78, 10.81, pp. 500--501]{Battin_1999}:
\begin{equation}
  \dot{L} = \frac{h}{r^2}+\frac{r}{h}F_h\tan\frac{i}{2}
  \sin(L-\Omega).\label{eq:dL}
\end{equation}
The time derivative of $\theta$ is obtained by differentiation of both
sides of equation
\[
\tan\theta = \frac{r\dot{r}c}{c^2-\mu r},
\]
which is a consequence of (\ref{eq:cth}), (\ref{eq:sth}). We
first use 
\begin{equation}
  \ddot{r}=-\frac{\mu}{r^2}+\frac{c^2}{r^3}-\frac{2\mathscr{U}}{r}+F_r,
  \label{eq:ddr}
\end{equation}
to get
\[
\dot{\theta} = \frac{c}{r^2}-\frac{c}{\mu rg}[(2\mathscr{U}-rF_r)\cos\theta+
  \tilde{\varsigma}\dot{c}\sin\theta].
\]
Then, we replace $\dot{c}$ with the expression in (\ref{eq:dc}) and find
\begin{equation}
\dot{\theta} = \frac{c}{r^2}-\frac{1}{\mu g}\Bigl[(\tilde{\varsigma}r\sin\theta)\dot{\mathscr{E}}+
  \Bigl(\frac{c}{r}\cos\theta+\tilde{\varsigma}\dot{r}\sin\theta\Bigr)(2\mathscr{U}-rF_r)\Bigr].
\label{eq:dth}
\end{equation}
From equations (\ref{eq:gdPsi}), (\ref{eq:dL}), (\ref{eq:dth}) we have
\begin{equation}
  g\dot{\Psi} = \frac{g}{r^2}(h-c)-\frac{gr}{h}\hat{w}_hF_h+
  \frac{\tilde{\varsigma}r}{\mu}\dot{\mathscr{E}}\sin\theta+\frac{1}{c}
    (\tilde{\varsigma}g+\varsigma\cos\theta)(2\mathscr{U}-rF_r),\label{eq:gdPsi_2}
\end{equation}
where (see \ref{eq:w_XYh})
\[
\hat{w}_h = -\tan\frac{i}{2}\sin(L-\Omega),
\]
and we applied the substitution
\[
\frac{c}{r}\cos\theta+\tilde{\varsigma}\dot{r}\sin\theta = \frac{\mu}{c}(\tilde{\varsigma}g+
\varsigma\cos\theta).
\]

The expressions of $\dot{g}$, $g\dot{\Psi}$, given in (\ref{eq:dg}),
(\ref{eq:gdPsi_2}), are inserted in (\ref{eq:dp1_2}), (\ref{eq:dp2_2}),
and by using the definitions of $p_1$, $p_2$ (see \ref{eq:p1},
\ref{eq:p2}) and the relation $L=\Psi+\theta$, we get equations
(\ref{eq:dp1}), (\ref{eq:dp2}).

\section{Time derivative of $\mathcal{L}$}
\label{sec:delldt}
From equation (\ref{eq:elle_2}) we have
\[
\dot{\mathcal{L}} = \dot{\mathcal{K}}(1-p_1\sin\mathcal{K}-
p_2\cos\mathcal{K})+\dot{p}_1\cos\mathcal{K}-
\dot{p}_2\sin\mathcal{K},
\]
which can be written as
\begin{equation}
  \dot{\mathcal{L}} = \dot{\mathcal{K}}\frac{r}{\textsl{\textrm{a}}}+\dot{p}_1\cos\mathcal{K}-
  \dot{p}_2\sin\mathcal{K},
  \label{eq:dL_2}
\end{equation}
taking into account relation (\ref{eq:r_2}). We first consider
$\dot{p}_1\cos\mathcal{K}-\dot{p}_2\sin\mathcal{K}$. Using
(\ref{eq:dp1}), (\ref{eq:dp2}), the following four terms will appear:
\[
p_1\sin\mathcal{K}+p_2\cos\mathcal{K},\qquad
p_1\cos\mathcal{K}-p_2\sin\mathcal{K},
\]
and
\begin{equation}
  \varsigma(\cos L\cos\mathcal{K}+\sin L\sin\mathcal{K}),\qquad
  \sin L\cos\mathcal{K}-\sin\mathcal{K}\sin L,
  \label{eq:csLK}
\end{equation}
where $\varsigma$ is introduced in (\ref{eq:aux}). The first two
terms are written as functions of $r$, $\dot{r}$,
$\textsl{\textrm{a}}$ through (\ref{eq:r_2}),
(\ref{eq:rrd_2}). Solving equations (\ref{eq:scL}) for
$\sin\mathcal{K}$, $\cos\mathcal{K}$ we get
\begin{align*}
  \sin\mathcal{K} & = \frac{1}{\beta}\Bigl[(1-\alpha p_1^2)
    \Bigl(\frac{r}{\textsl{\textrm{a}}}\sin L+p_1\Bigr)-\alpha p_1p_2
    \Bigl(\frac{r}{\textsl{\textrm{a}}}\cos L+p_2\Bigr)\Bigr],\\
  \cos\mathcal{K} & = \frac{1}{\beta}\Bigl[(1-\alpha p_2^2)
    \Bigl(\frac{r}{\textsl{\textrm{a}}}\cos L+p_2\Bigr)-\alpha p_1p_2
    \Bigl(\frac{r}{\textsl{\textrm{a}}}\sin L+p_1\Bigr)\Bigr],
\end{align*}
where $\alpha$, $\beta$ are defined in (\ref{eq:alpha}), (\ref{eq:bega}). In
particular, we used the relation
\[
\beta = 1-\alpha(p_1^2+p_2^2).
\]
The expressions above for $\sin{\mathcal{K}}$, $\cos{\mathcal{K}}$ are
needed to write the two terms (\ref{eq:csLK}) as
\begin{align*}
  \varsigma(\cos L\cos\mathcal{K}+\sin L\sin\mathcal{K}) & =
  1-\varsigma+\beta\left[\varsigma^2-\alpha(1-\varsigma)^2\right],\\
  \sin L\cos\mathcal{K}-\cos L\sin\mathcal{K} & = \frac{c\dot{r}}{\mu}
  \left[\varsigma+\alpha(1-\varsigma)\right],
\end{align*}
where we noticed that
\begin{align}
  p_1\sin L+p_2\cos L & = \frac{1}{\varsigma}-1,\nonumber\\
  p_2\sin L-p_1\cos L & = \frac{c\dot{r}}{\mu}\label{eq:aux},
\end{align}
and
\begin{equation}
  \alpha\beta = 1-\alpha,\qquad
  \frac{r}{\textsl{\textrm{a}}} = \beta^2\varsigma.\label{eq:aux_2}
\end{equation}
We can write
\begin{align}
  & \dot{p}_1\cos\mathcal{K}-\dot{p}_2\sin\mathcal{K} =\nonumber \\
  & \biggl(\frac{h-c}{r^2}-\frac{r}{h}\hat{w}_hF_h\biggr)\Bigl(1-\frac{r}{\textsl{\textrm{a}}}\Bigr)
  +\frac{1}{c}\biggl[\Bigl(1-\frac{r}{\textsl{\textrm{a}}}\Bigr)(1+\alpha\beta\varsigma)\nonumber \\
  &  +\alpha(1+\beta\varsigma)-\frac{(r\dot{r})^2}{c\mu}\textsl{\textrm{w}}\biggr](2\mathscr{U}-rF_r)+
  \frac{r\dot{r}}{\mu^2}c\Bigl[\alpha+\varsigma\Bigl(1-\frac{r}{\textsl{\textrm{a}}}\alpha\Bigr)\Bigr]
  \dot{\mathscr{E}},
  \label{eq:dp12K}
\end{align}
where $\textsl{\textrm{w}}$ is introduced in (\ref{eq:w}) and we
applied the relations
\begin{align*}
  & \tilde{\varsigma}\Bigl(1-\frac{r}{\textsl{\textrm{a}}}\Bigr)+1-\varsigma+\beta\left[\varsigma^2-
    \alpha(1-\varsigma)^2\right] =
  \Bigl(1-\frac{r}{\textsl{\textrm{a}}}\Bigr)(1+\alpha\beta\varsigma)+\alpha(1+\beta\varsigma),\\
  & \tilde{\varsigma}c\left[\varsigma+\alpha(1-\varsigma)\right]-r\varsigma \textsl{\textrm{w}}= c
  \Bigl[\alpha+\varsigma\Bigl(1-\frac{r}{\textsl{\textrm{a}}}\alpha\Bigr)\Bigr].
\end{align*}

We deal with the time derivative of $\mathcal{K}$. From
(\ref{eq:Psi}), (\ref{eq:K}) we have
\begin{equation}
  \dot{\mathcal{K}} = \dot{L}+\dot{G}-\dot{\theta}.
  \label{eq:dK}
\end{equation}
The expressions of $\dot{L}$, $\dot{\theta}$ are shown in
(\ref{eq:dL}), (\ref{eq:dth}). By differentiation of both sides of
equation
\[
\tan G = \frac{r\dot{r}}{\textsl{\textrm{w}}(\textsl{\textrm{a}}-r)},
\]
which follows from (\ref{eq:r}), (\ref{eq:rrd}), and using
(\ref{eq:ddr}), we find
\[
\dot{G} = \frac{\textsl{\textrm{w}}}{r}-\frac{1}{g\sqrt{\mu\textsl{\textrm{a}}}}\Bigl[\frac{\mu r\sin\theta}
  {2c\textsl{\textrm{a}}^2}(r+\textsl{\textrm{a}})\dot{\textsl{\textrm{a}}}+(\cos G)(2\mathscr{U}-rF_r)\Bigr].
\]
Then, considering that
\[
\cos G = \frac{\cos\theta + g}{1 + g\cos\theta},\qquad
\dot{\textsl{\textrm{a}}} = \frac{2\textsl{\textrm{a}}^2}{\mu}\dot{\mathscr{E}},
\]
we can write
\begin{equation}
  \dot{G} = \frac{\textsl{\textrm{w}}}{r}-\frac{r}{cg\sqrt{\mu\textsl{\textrm{a}}}}\Bigl[\sin\theta(r+\textsl{\textrm{a}})
    \dot{\mathscr{E}}+\frac{\mu}{c}(\cos\theta+g)(2\mathscr{U}-rF_r)\Bigr].
  \label{eq:dG}
\end{equation}
From equations (\ref{eq:dK}) and (\ref{eq:dL}), (\ref{eq:dth}),
(\ref{eq:dG}) we obtain
\begin{align}
  \dot{\mathcal{K}} & = \frac{\textsl{\textrm{w}}}{r}+\frac{h-c}{r^2}-\frac{r}{h}\hat{w}_hF_h
  +\frac{1}{c}\Bigl[1+\alpha\Bigl(1-\frac{r}{\textsl{\textrm{a}}}\Bigr)\Bigr](2\mathscr{U}-F_rr)
  \nonumber\\
  & \quad\, -\frac{r\dot{r}\alpha}{\mu}\Bigl(\frac{1}{\textsl{\textrm{w}}}-
  \frac{r}{c}\Bigr)\dot{\mathscr{E}}.
  \label{eq:dK_2}
\end{align}

After using (\ref{eq:dp12K}), (\ref{eq:dK_2}) in equation
(\ref{eq:dL_2}) we obtain
\begin{align*}
  \dot{\mathcal{L}} & = \nu+\frac{h-c}{r^2}-\frac{r}{h}
  \hat{w}_hF_h+\Bigl(\frac{r\dot{r}c}{\mu^2}\tilde{\varsigma}\alpha\Bigr)\dot{\mathscr{E}}
  +\frac{1}{c}\biggl[(1+\alpha)(1+\beta\varsigma)\nonumber\\
    & \quad\, -\beta\varsigma\frac{r}{\textsl{\textrm{a}}}
    -\frac{(r\dot{r})^2}{c\mu}\textsl{\textrm{w}}\biggr](2\mathscr{U}-rF_r).
  \label{eq:dell}
\end{align*}
Finally, the expression of $\dot{\mathcal{L}}$ reported in (\ref{eq:dell})
follows noting that
\[
(1+\alpha)(1+\beta\varsigma)-\beta\varsigma\frac{r}{\textsl{\textrm{a}}}-\frac{(r\dot{r})^2}{c\mu}\textsl{\textrm{w}}=
\frac{1}{\alpha}+\alpha\Bigl(1-\frac{r}{\textsl{\textrm{a}}}\Bigr),
\]
where we used relations (\ref{eq:aux}), (\ref{eq:aux_2}), and
\[
(p_2\sin L-p_1\cos L)^2=\frac{2\varsigma-1}{\varsigma^2}-\beta^2. 
\]

\bibliographystyle{spbasic}
\bibliography{mybib}

\end{document}